\documentclass[12pt]{article}

\usepackage{graphicx}
\usepackage{times}
\usepackage{tikz}
\usepackage{cite}
\usepackage{float}
\usepackage{siunitx}  
\DeclareSIUnit\molar{M} 

\newenvironment{sciabstract}{%
\begin{quote} \bf}
{\end{quote}}

\title{Pinned Boundaries Delay Contraction and Shape Stress Relaxation in Active Gels}

\author
{Aniket Marne$^{1,2}$,James Clarke$^{1}$, Aravind Rao $^{3}$, Hyunjae Lee$^{1}$, Kyla Wong$^{1}$,\\ Aditya Sriram$^{4}$, Rae Robertson-Anderson$^{4}$, Moumita Das$^{3}$,Jos\'{e} Alvarado$^{1\ast}$\\
\\
\normalsize{$^{1}$Center for Nonlinear Dynamics, Department of Physics,} \\[0.5ex] 
\normalsize{$^{2}$Texas Materials Institute,} \\[0.5ex]
\normalsize{The University of Texas at Austin, Austin, TX, USA}\\[0.5ex]
\normalsize{$^{3}$School of Physics and Astronomy,} \\[0.5ex]
\normalsize{Rochester Institute of Technology, Rochester, NY, USA}\\[0.5ex]
\normalsize{$^{4}$Department of Physics and Biophysics,} \\[0.5ex]
\normalsize{University of San Diego, San Diego, California 92110}\\\\
\normalsize{$^\ast$ : To whom correspondence should be addressed; E-mail:  alv@chaos.utexas.edu.}
}

\date{}

\begin{document} 



\baselineskip24pt


\maketitle

\begin{sciabstract}
Cells dynamically generate, transmit, and dissipate stress.
Central to these processes is the actomyosin cortex, an active contractile material that powers many instances of cellular mechanical behavior.
While prior studies have focused on freely contracting actomyosin systems, the role of mechanical constraints, such as adhesion to boundaries, remains less explored.
To this end, we employ reconstituted actomyosin gels to dissect the physical principles underlying cellular contractility.
We investigate the contraction dynamics of actomyosin active gels under pinned boundary conditions, where the gel is adhered transversely to two opposing surfaces, mimicking the supracellular actomyosin network of tissues and embryos.
We observe that pinned contraction results in the build-up of stress, which delays contraction, produces intermittent dynamics, and introduces spatial non-uniformity in the strain field.
Stress is relieved by several distinct pathways, including
slow, active-stress-driven, symmetric constriction;
and rapid, defect-driven mechanisms such as detachment from boundaries and internal rupture.
We further develop a hydrodynamic model that incorporates elastic, viscous, and active stress terms.
It distinguishes between stress‑accumulation and stress‑release phases and connects changes in active stress to observed intermittent contraction dynamics.
The model also reveals distinct energy relaxation rates before and after detachment events, providing insight into the temporal modulation of stress dissipation.
Finally, we compare our experiment and model with numerical simulations, which confirm our observations.
In addition, simulations reveal how internal energy is generated and dissipated over the course of stress build-up and relaxation.
Our findings underscore the importance of boundary conditions and spatial heterogeneity in internal stress, which together shape the mechanical behavior of contractile active gels.
These results have implications for understanding stress regulation in cellular and tissue-scale contexts and may inform the design of adaptive soft materials and bioinspired robotics capable of interacting with their mechanical environment.
\end{sciabstract}

\section*{Introduction}

Although individual cells can function autonomously, entire collections of cells often work in concert to structure and remodel tissues and embryos\cite{sutherlandPulsedActomyosinContractions2020}. Specific instances of such collective remodeling include apical constriction \cite{dehapiotAssemblyPersistentApical2019}, convergent extension, folding, tension homeostasis, epiboly, tissue closure, and wound healing. These processes rely on coordinated mechanical responses from ensembles of autonomous cellular agents - a phenomenon that remains a central topic in developmental biology and tissue mechanics.  Two key cellular components drive these coordinated behaviors: the actomyosin cytoskeleton and cell adhesions. Together, they form a dynamic supracellular actomyosin network capable of generating tissue-level forces and shape changes. While much attention has been given to the role of intracellular contractility and intercellular adhesion, the influence of boundary adhesion, the mechanical interactions at tissue interfaces or edges, has received comparatively less focus.  

The actomyosin cytoskeleton is a central driver of mechanical behavior at the single-cell level, powering shape changes, motility, and intracellular transport \cite{mitchisonActinBasedCellMotility1996}. At the molecular scale, the spatial arrangement of actin and myosin \cite{murrellForcingCellsShape2015} along with the architecture of actin structures \cite{koenderinkArchitectureShapesContractility2018} gives rise to diverse contractile behaviors. These forces are transmitted across cells through junctional and adhesion molecules, which link the cytoskeleton to neighboring cells  \cite{arnoldRhoGTPasesActomyosin2017, nelsonRegulationCellCell2008,leckbandCadherinAdhesionMechanotransduction2014} and to the extracellular matrix \cite{humphreyMechanotransductionExtracellularMatrix2014a, kechagiaIntegrinsBiomechanicalSensors2019}. The actomyosin cytoskeleton and adhesion molecules work together to establish supracellular contractile structures, such as cables and rings, which generate contractile stresses. Furthermore, to generate mechanical work and drive tissue-level shape change, adhesion molecules also connect supracellular contractile structures to the surrounding tissue.

A key question in tissue mechanics is how contractility and adhesion cooperate to drive large-scale remodeling.  \emph{In vitro} assays of reconstituted actomyosin networks have provided insights into this interplay. Earlier studies focused on actomyosin contractility in the absence of adhesion to exterior surfaces (“free contraction”) \cite{alvaradoMolecularMotorsRobustly2013}. When unanchored to exterior boundaries, actomyosin networks often retain their original shape when deforming. These kinds of contraction events mimic volumetric intracellular contraction events, such as in chromosome congression in starfish oocyte\cite{lenartContractileNuclearActin2005}. However, more complex behavior occurs when contractile actomyosin interacts with boundaries \cite{alvaradoForcePercolationContractile2017}. Inspired by the actomyosin cortex, studies have anchored planar actin networks to lipid bilayers, either in a planar configuration \cite{murrellFactinBucklingCoordinates2012} or in liposomes \cite{carvalhoCellsizedLiposomesReveal2013, sakamotoActiveTensionMembrane2024}. As a region of actomyosin generates contractile stress, it interacts mechanically with its passive environment \cite{linsmeierDisorderedActomyosinNetworks2016}. Anchoring networks at opposing sides (“pinned boundary conditions”) allows the gel to bear tension for sustained periods of time. An early study leveraged this condition to estimate contractile force  \cite{bendixQuantitativeAnalysisContractility2008a}. Light scattering experiments revealed avalanche-like decorrelation events that precede depinning in gels that store mechanical stress \cite{alvaradoUncoveringDynamicPrecursors2019}.

Beyond stress generation, stress relaxation is essential for mechanical integrity. Cytoskeletal fluidization under transient stretch \cite{trepatUniversalPhysicalResponses2007,kimDeterminantsFluidlikeBehavior2014} and actomyosin-dependent relaxation in epithelial tissues highlight the importance of dynamic stress redistribution \cite{khalilgharibiStressRelaxationEpithelial2019}. Without the ability to distribute stresses across multiple cells via mechanical coupling, tissues would be susceptible to failure modes such as fracture \cite{wyattQuestionTimeTissue2016}. Yet, how contractile gels accumulate, store, and relax stress under boundary constraints remains poorly understood. 

In this study, we investigate the mechanical behavior of \textit{in vitro} reconstituted actomyosin active gels under pinned boundary conditions. To mimic cellular adhesion, we anchor the gels to two opposing boundaries, enforcing Dirichlet boundary conditions ($U = 0$) along parallel planes. Using fluorescence microscopy, we track the temporal evolution of the gels and apply image-based strain tensor analysis to quantify deformation. 
Our experiments reveal that boundary adhesion significantly delays macroscopic contraction compared to free-contraction controls, resulting in a logistic-like strain evolution. Spatial mapping of the strain field shows pronounced inhomogeneities in both strain and inferred stress distributions, highlighting the role of boundary constraints in shaping mechanical behavior. Additionally, we observe that the orientation of principal strains becomes more heterogeneous under pinned conditions, suggesting disrupted symmetry and anisotropic force transmission. To further understand these dynamics, we compare overall contractility between free and boundary-anchored gels. The presence of adhesion not only slows contraction but also alters the spatial organization of strain, indicating that boundary conditions modulate both the magnitude and directionality of mechanical responses.
To interpret these findings, we develop numerical simulations that replicate the experimental setup and validate the observed behaviors. We also employ a simplified hydrodynamic model to estimate internal stress and characterize the relaxation timescale. This model reveals that in the absence of global contraction, due to boundary adhesion, the gel undergoes internal reorganization, which governs stress relaxation and mechanical equilibration. Finally, we report various avenues by which actomyosin gels relax stored stresses.

\section*{Results}

We investigate relaxation after tensile stress build up for \textit{in vitro} actomyosin gels subject to transverse anchoring. To achieve this, actomyosin gels were attached to opposite walls to ensure Dirichlet boundary conditions with displacement field $U=0$ at the left and right side of the chamber (see Methods). Contractile gel samples are prepared, and imaged via an epifluorescence microscope (see Methods). We conducted $N=19$ experiments out of which only $N=12$ were considered appropriate for analysis (see Methods). Of these twelve analyzed experiments, $N=3$ are control samples where boundaries were passivated to ensure that the gels contracted freely. From the remaining $N=9$ pinned experiments, we select $N=5$ samples that exhibited robust adhesion along both opposing boundaries for some finite duration, denoted as $\tau_D$. In these “detaching” samples, the gel remained anchored and deformed slightly before breaking free and fully contracting. In parallel, we model pinned actomyosin contraction by modeling the actin network using a triangular lattice in a similar approach to previous studies (see Methods). \cite{broedersz2011molecular, kumar2023range, lee2021active}. Finally we consider the remaining $N=4$ pinned samples that exhibit a variety of contraction behaviors, which do not fit in the “detaching” category.


\subsection*{Pinned boundaries during active contraction permits stress build-up}

\begin{figure}[H]
    \centering
\includegraphics[width=\linewidth]{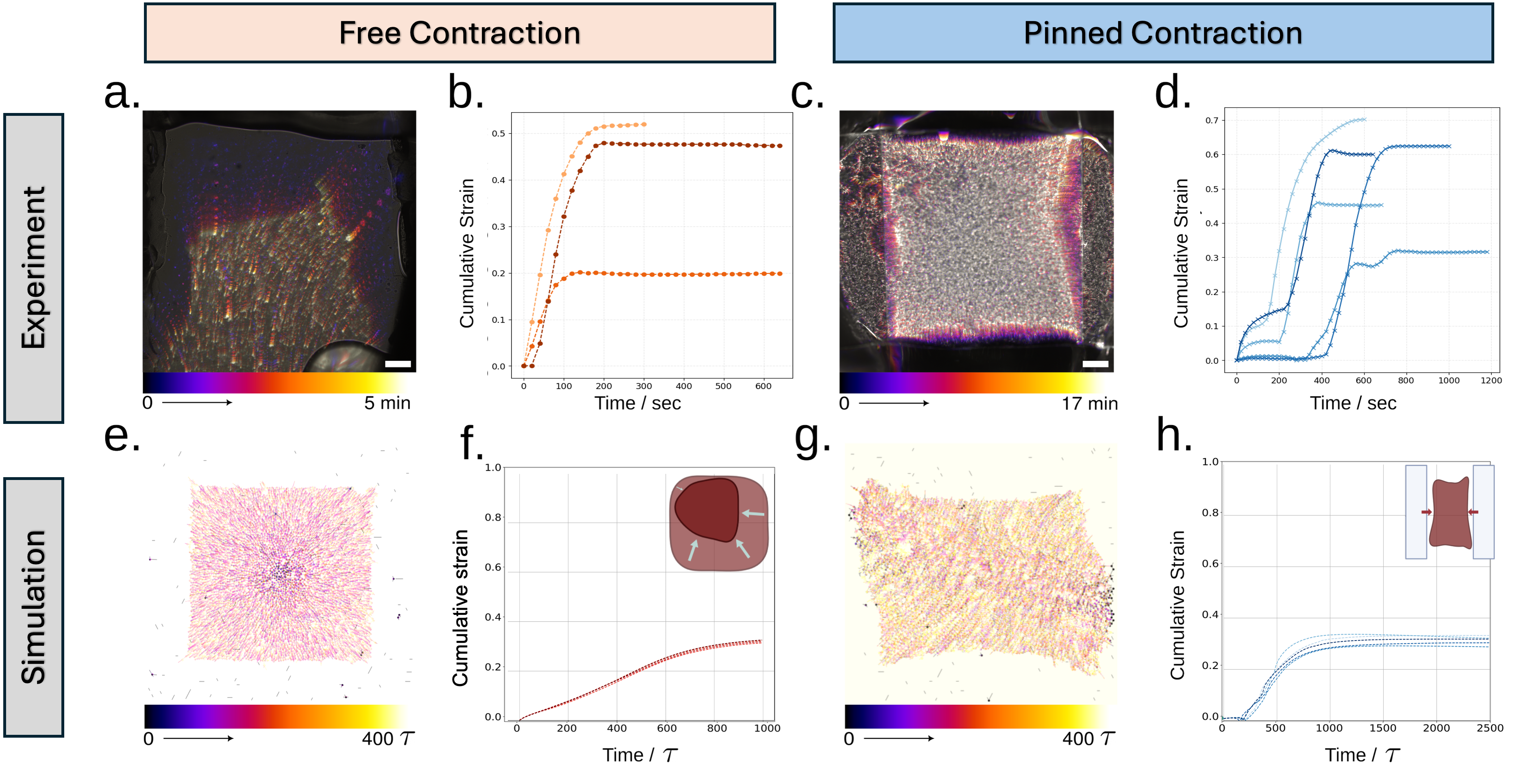}
\caption{Comparison between strain responses between experiments and simulations for both free and pinned contractions. (a) Time overlay for representative experiment of free contraction - our control population ($N=3$), (b) Cumulative Hencky strain for $N=3$ control samples. (C) Time overlay for representative experiment of pinned experiments - with a detach relaxation pathway ($N=5$), (d) Cumulative Hencky strain for $N=5$ pinned samples with a detach relaxation pathway.For panels b, and d data is shown for tensorial Hencky strain determined using PIV  methodology.(e) Time overlay for   Control (free contraction) simulation of representative simulated network ($N=5$).     (f) Cumulative Hencky strain for $N=5$ simulated free contraction experiments.(g) Time overlay for pinned simulation of  representative simulated network ($N=5$).     (h) Cumulative Hencky strain for $N=5$ simulated pinned contraction experiments.}
    \label{fig_results_strain}
\end{figure}

To ensure consistency across samples, all gels were prepared using Nitrophenyl Ethyl (NPE) caged-ATP, which enabled controlled initiation of contraction via UV-triggered ATP release~\cite{clarke2025nonlinearcontractileresponseactomyosin}. For this study, we focused exclusively on the first contraction event to avoid confounding effects from repeated activation cycles (see Methods). 
We begin by examining the behavior of gels undergoing \textit{unconstrained free contraction}, which serve as our control population ($N=3$). A representative dataset is shown in Fig.~\ref{fig_results_strain}a. As the gel contracts, we observe a smooth and spatially uniform reduction in size, consistent with expectations for freely relaxing networks. To investigate the global contractile properties of a given gel, epifluorescence images were acquired and processed (see Methods) to give a measure of the boundary Hencky strain based on the total area change of the gel. Along with that, to investigate the spatially resolved strain changes, a particle image velocimetry (PIV) technique was used (see Methods) to track local displacements, and then determine the strain-tensor field within the gel\cite{choudhary_investigating_2026}. 

We analyze the displacement field $u(X,t)$ obtained from PIV, where $X = (x, y)$ denotes spatial position and $t$ is time. From this displacement field, we compute the deformation gradient tensor $F(X) = I + \nabla u$, where $I$ is the identity tensor. $F(X)$ describes how an infinitesimal material element is locally stretched, sheared, and rotated.  Using $F$, we calculate the Green-Lagrange strain tensor $\epsilon(X) = \frac{1}{2}(F^T F - I)$. This quantity has two advantages. First, it remains accurate for large strains, where linear strain measures fail to capture geometric nonlinearities. Second, it gives us the tensorial strain at each location in the gel. Using the spatial strain field, we compute the Hencky strain (see Methods)(Fig.~\ref{fig_results_strain}b,d) to quantify the temporal evolution of deformation in both free‑ and pinned‑contraction experiments. Here , we have plotted absolute value of contractile strain. In free‑contraction samples, the cumulative Hencky strain increases monotonically from the onset of the experiment across all control gels ($N=3$). Notably, strain accumulation begins immediately at $t=0$, coincident with ATP availability, and continues smoothly throughout the contraction process (Fig.~\ref{fig_results_strain}b). The strain growth follows an approximately exponential rise before reaching a plateau at long times.

Pinned‑contraction experiments display a distinctly different strain evolution. In representative datasets (Fig.~\ref{fig_results_strain}c), the gel exhibits minimal macroscopic motion during the initial period following ATP addition. Nevertheless, the Hencky strain increases steadily during this early interval, indicating ongoing internal deformation (SI Fig. 6; strain field snapshot in lag phase). After a characteristic stall time (approximately the first five minutes in Fig.~\ref{fig_results_strain}d), the gel detaches from the boundary and undergoes rapid contraction, accompanied by a sharp increase in cumulative strain. 
The Hencky strain eventually saturates at late times (Fig.~\ref{fig_results_strain}d). This behavior is observed consistently across pinned samples exhibiting detachment ($N=5$).

Comparing the two contraction modes reveals clear differences in the timing and structure of strain accumulation. Free‑contraction gels exhibit smooth, uninterrupted strain growth beginning at $t=0$, with no detectable lag phase. In contrast, pinned samples display a two‑stage strain evolution: an initial period of slow but continuous strain buildup, followed by a rapid increase after detachment. Despite these differences in early‑ and intermediate‑time dynamics, both free and pinned contractions ultimately reach similar levels of cumulative strain, indicating comparable final degrees of compaction.
We interpret this observation in the following way. The mechanical constraints placed on the gel prevent it from contracting immediately. However, it is accumulating internal, active stresses. Once the internal stress exceeds a threshold value set by the adhesion strength, the gel begins to fully contract.

Simulation results reproduce these experimental trends. Free‑contraction simulations (Fig.~\ref{fig_results_strain}e–f) exhibit immediate strain accumulation with an exponential rise and no lag, closely mirroring experimental observations. Pinned‑contraction simulations (Fig.~\ref{fig_results_strain}g–h) similarly capture the delayed onset of rapid contraction following a period of constrained strain growth, followed by saturation at long times.

The immediate rise in Hencky strain observed in free‑contraction experiments indicates that, in the absence of external constraints, active stresses generated within the gel are directly converted into macroscopic deformation. The exponential form of strain growth suggests a contractile process governed by internal activity rather than external resistance.

In pinned‑contraction samples, the gradual increase in strain during the early, adhered phase reflects the accumulation of internal active stresses under mechanical constraint. Although macroscopic contraction is suppressed, internal remodeling and deformation proceed within the gel network. Detachment occurs once these stresses exceed a threshold set by boundary adhesion, triggering rapid contraction and a corresponding surge in strain. The subsequent saturation reflects the system reaching a mechanically relaxed, fully compacted state.

The strong qualitative agreement between experiments and simulations under both free and pinned conditions demonstrates that the model accurately captures how internal contractility and boundary constraints together shape strain evolution, including the emergence of lag‑phase behavior and rapid contraction upon release of constraints.

The strain analyses in Figure ~\ref{fig_results_strain}a–h reveal not only the mechanical signatures of free and pinned contractions but also their important biological implications. The early-time rise in tensorial Hencky strain, even when boundary strain remains unchanged, indicates that cells generate and transmit internal contractile stresses well before macroscopic contraction becomes visible. This suggests that intracellular actomyosin networks and cell–cell junctions begin reorganizing and loading the matrix long before tissue‑scale deformation occurs. The heterogeneous strain patterns observed even in free contraction further imply that contractile forces are not uniformly distributed, pointing to spatially variable cell activity, differences in local cytoskeletal tension, or heterogeneous connectivity in the extracellular network. In pinned samples, the prolonged mismatch between tensorial and boundary strains underscores the role of adhesive constraints in storing internal elastic stress, which is later released abruptly upon detachment-resembling biological processes such as adhesion‑mediated force buildup, mechanosensing, and sudden cell–matrix rupture events. Together, these observations indicate that the gel does not behave as a passive contracting material but instead reflects active, spatially patterned force generation. This motivates the development of a mechanistic stress model capable of linking local strain buildup, boundary adhesion, and active force generation into a unified framework. We therefore next introduce our continuum‑level active stress model, which captures how intrinsic biological contractility translates into the observed strain dynamics across free and pinned conditions.

To interpret the strain dynamics described above and connect them to the underlying force ‑ generation mechanisms, we model the actomyosin gel as an active viscoelastic material. Such materials exhibit three key behaviors: Elastic deformation representing reversible filament stretching,Viscous dissipation arising from crosslink unbinding and filament rearrangement, and Active stress generation driven by ATP‑powered motor activity.

We therefore express the total Cauchy stress tensor as the sum of elastic, viscous, and active contributions:
$\sigma = \sigma_e + \sigma_d + \sigma_a$
The elastic response is modeled using a linear constitutive law,
$\sigma_e = G\,\epsilon$,
where $G$ denotes the effective shear modulus and $\epsilon$ is the Green–Lagrange strain tensor derived from PIV measurements.
The viscous response captures time‑dependent relaxation of the network,
$\sigma_d = \eta\,\dot{\epsilon}$,
where $\eta$ is the effective viscosity and $\dot{\epsilon}$ is the strain‑rate tensor.
The active stress arises from myosin‑driven contractility and the consumption of ATP. To model the gradual reduction in motor activity as ATP becomes depleted, we adopt a time‑dependent active term of the form:
$\sigma_a = \zeta e^{-\beta t} \epsilon$, where $\zeta$ is a material parameter reflecting motor density and ATP sensitivity, and $\beta$ characterizes the rate of ATP depletion.
Thus, the total stress is given by:
$\sigma = G \epsilon + \eta \dot{\epsilon} + \zeta e^{-\beta t} \epsilon$.

This formulation allows us to estimate internal stress fields directly from PIV‑derived strain measurements, providing a framework for interpreting how internal contractile forces accumulate and dissipate over time. Notably, the decaying active term captures the progressive reduction of contractile force generation as biochemical energy sources diminish.
Importantly, this model also clarifies the behaviors observed experimentally. In pinned contractions, the boundary constraints prevent immediate macroscopic shortening, allowing internal elastic and active stresses to accumulate which is consistent with the divergence of boundary and tensorial strain in early time points and the sigmoidal dynamics observed upon detachment. In free contractions, where no boundary forces oppose motion, stress and strain evolve in closer synchrony, though local spatial heterogeneity still leads to small but reproducible differences between strain measures.
Overall, this active viscoelastic framework provides a mechanistic basis for linking local strain evolution to global stress dynamics, enabling quantitative interpretation of the experimental strain behaviors described in Figure ~\ref{fig_results_strain}.

To further characterize the mechanical behavior of contracting gels, we introduce an activity parameter $\mathcal{E}$, which estimates the instantaneous mechanical energy density generated per unit time. This parameter is defined using the stress field obtained from frame‑by‑frame strain measurements, and scales as $\mathcal{E} \sim \sigma^{2}$,
reflecting the idea that active motor‑generated stress directly contributes to the mechanical work performed on the network.

\begin{figure}[H]
    \centering
    \includegraphics[width=0.95\linewidth]{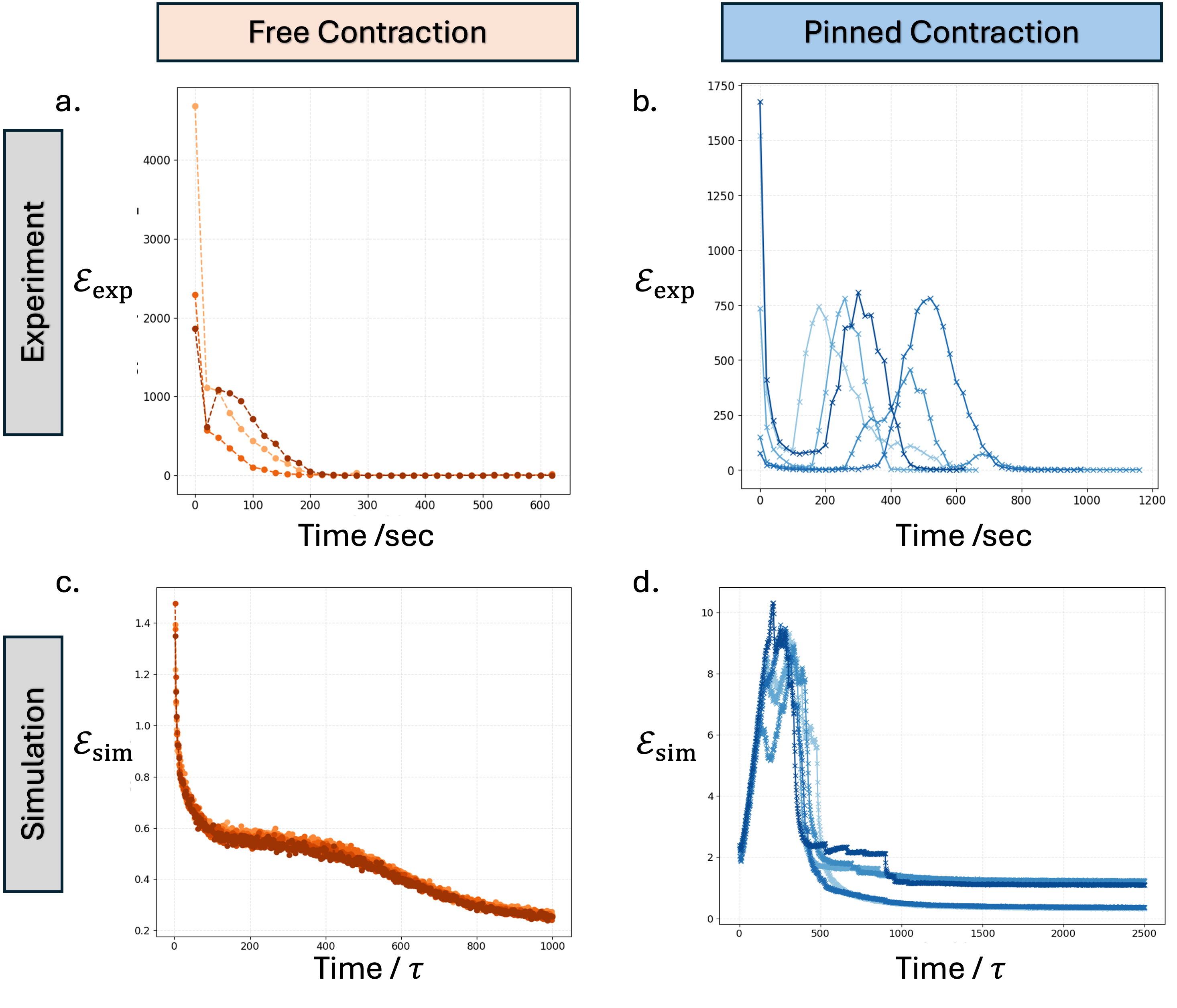}
\caption{Energy density changes shows relaxation via local structural rearrangement within gel and detachment (a) Energy density contribution change based on a model, $\mathcal{E}$, evolution with parameters set to $G= 1$ , $\eta=250$ , $\zeta = 10$, and $\beta = 0.01$ units with ATP consumption in control (circles) samples with free contraction  and (b)detach samples (x's) with pinned contraction.  (c) Energy density contribution change, $\mathcal{E}$, evolution resulting from simulations for free contraction simulation (control) and (d) pinned contraction simulation. Units are arbitrary based on parameters used for model and simulation.}
\label{fig_results_energy}
\end{figure}

An activity parameter $\mathcal{E}$ captures the combined effects of active stress generation, viscoelastic relaxation, and boundary interactions during contraction. Figure~\ref{fig_results_energy} compares the temporal evolution of this parameter for both experiments and simulations under free and pinned boundary conditions, enabling direct comparison between unconstrained relaxation and mechanically constrained dynamics.
For the experimental measurements, the activity parameter $\mathcal{E}_{exp}$ is computed from PIV-derived displacement fields obtained during contraction. In the free-contraction control experiments (Fig.~\ref{fig_results_energy}a), $\mathcal{E}_{exp}$ evolves smoothly and continuously throughout the contraction process. This gradual decay is consistent with steady viscoelastic relaxation and progressive strain redistribution in the absence of external mechanical constraints. The absence of abrupt transitions indicates that stress relaxation occurs continuously through distributed network remodeling rather than through discrete failure events.
In contrast, pinned-contraction experiments (Fig.~\ref{fig_results_energy}b) exhibit markedly different energetic dynamics due to boundary confinement. Prior to detachment, $\mathcal{E}_{exp}$ undergoes an initial decay phase even though the global strain remains near zero. This behavior demonstrates that the network continues to dissipate internally generated stresses through local filament rearrangements and viscoelastic relaxation despite macroscopic suppression of contraction. At the moment of boundary detachment, $\mathcal{E}_{exp}$ displays a sharp, discontinuous increase, reaching a pronounced peak as accumulated elastic stresses are rapidly released. Following detachment, the activity parameter decays again as the gel transitions into an unconstrained relaxation regime similar to that observed in the control samples. All pinned samples exhibit qualitatively similar energetic signatures surrounding detachment, highlighting the reproducibility of the stress-release dynamics associated with boundary failure.

To compare these experimental observations with theory, we compute a corresponding simulation-based activity metric, $\mathcal{E}_{sim}$, by summing the total spring and bending energies of the network at each simulation timestep. In the free-contraction simulations (Fig.~\ref{fig_results_energy}c), $\mathcal{E}_{sim}$ exhibits a smooth relaxation profile analogous to the experimental control samples, reflecting gradual release of internally stored elastic energy during unconstrained contraction. This agreement indicates that the simulations successfully capture the dominant mechanics governing stress relaxation in freely contracting active gels.
Under pinned boundary conditions (Fig.~\ref{fig_results_energy}d), the simulations reproduce the principal energetic features observed experimentally. In particular, $\mathcal{E}_{sim}$ exhibits a localized peak centered on the detachment event, followed by rapid decay as the network relaxes into its post-detachment configuration. However, unlike the experiments, the simulated traces contain comparatively few data points prior to detachment and show limited pre-detachment evolution in the activity parameter. This difference arises because the simulated network is maintained under an idealized static pinned condition with minimal internal remodeling before failure, leading to relatively constant stored elastic energy until detachment occurs. By contrast, experimental actomyosin networks continue to undergo slow structural rearrangements and stress redistribution even while macroscopically constrained, producing the gradual pre-detachment decay observed in $\mathcal{E}_{exp}$. Thus, the discrepancy in pre-detachment temporal evolution reflects differences in remodeling dynamics and relaxation timescales between the physical and simulated systems rather than disagreement in the underlying mechanical response.
Taken together, the activity parameter $\mathcal{E}$ serves as a compact energetic descriptor that links active force generation, viscoelastic dissipation, and boundary-mediated stress release, complementing the strain- and stress-based analyses presented above.

Together, these experimental and simulation results reinforce the conclusion that boundary constraints strongly modulate active stress buildup and its eventual release. The differences between free and pinned contraction which are evident both in strain evolution (Fig.~\ref{fig_results_strain}) and in the rate and timing of energy changes (Fig.~\ref{fig_results_energy}) highlight the role of adhesion in governing how active stress accumulates, dissipates, and is redistributed across the gel. An additional observation, visible in Supplement Fig. 4 d, is that globally measured (boundary‑derived) strains are generally larger in magnitude than locally measured (tensorial) strains. One possible explanation is that local deformations within the gel include components misaligned with the principal strain direction, reducing the apparent magnitude when quantified locally but still contributing to the global area‑based measure. A similar discrepancy is observed in the free‑contraction controls (Supplement Fig.4 b), suggesting that such misalignment or non‑affine deformation is an inherent feature of the actomyosin network rather than a consequence of boundary pinning.
These observations motivate a more detailed investigation into strain non‑uniformity within the contracting gels. In the following section, we quantify the spatial structure of local strain fields to quantify non-uniformity in strain measurements.

\subsection*{Pinned boundaries induce spatial heterogeneity in strain}
\begin{figure}[H]
    \centering
    \includegraphics[width=\linewidth]{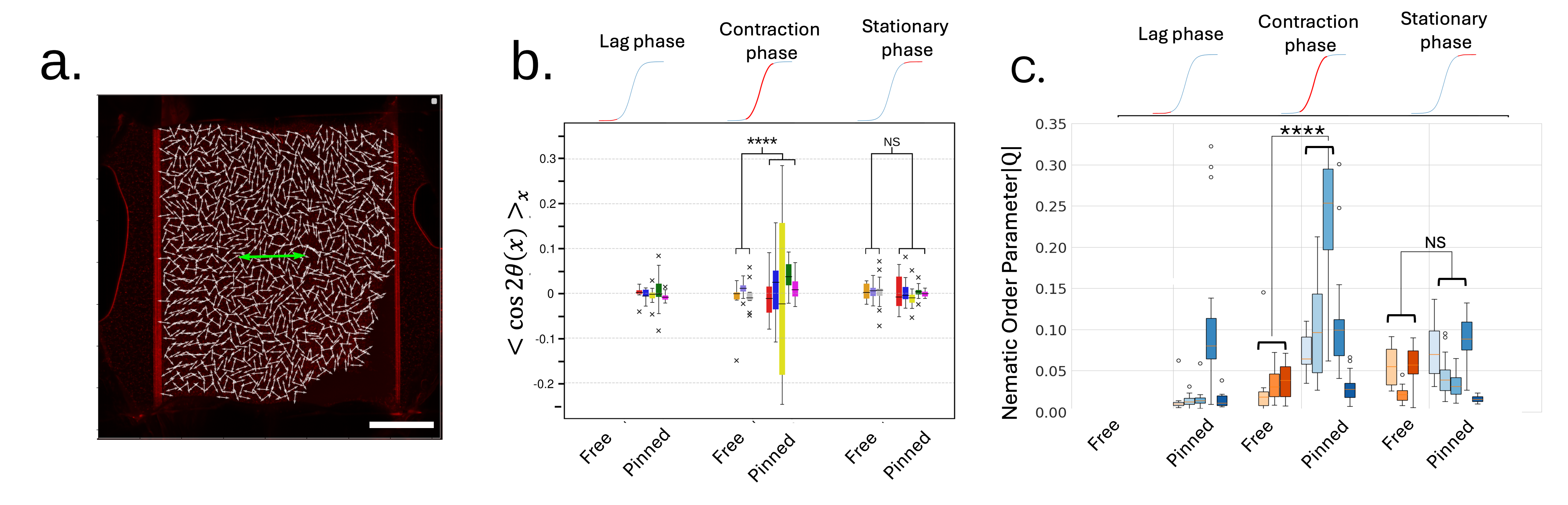}
\caption{Principal strain direction defines nematic order. (a) Representative visualization of principal strain directions overlaid on a epifluorescence micrograph of contractile gel. The white arrows represent local variations in strain direction $2\theta_p$. The green arrow represents the global principal strain direction $\langle 2\theta_p \rangle$, computed as the mean strain orientation across the entire region. (b) Mean of $\cos 2\theta$ distributions split on contraction region, type (control vs. pinned), and sample. (c) Nematic Order parameter $|Q|$distributions split on contraction region, type (control vs. pinned), and sample. To distinguish the contraction region, graphical representations of detaching pinned contraction evolutions are shown above the distributions with red overlay to distinguish the region of interest. The region on the left compares the two while gel is pinned. The region in the center compares the two while the gel contracts. The region on the right compares the two while gel contraction plateaus. Levene’s Tests are performed on region B, and region C for both parameters. We report a statistically significant difference (p=0.0001) and no statistically significant difference, respectively for $\langle \cos(2\theta) \rangle$. $|Q|$ shows statistically significant difference (p=0.00001) and no statistically significant difference, respectively for region B and C. T-Test on  $|Q|$ shows statistically significant difference (p=0.00001) and no statistically significant difference, respectively for region B and C.}
    \label{fig_results_director}
\end{figure}

The accumulation of stored contractile stresses in pinned actomyosin gels naturally raises the question of how these stresses are redistributed throughout the network during contraction. While the previous section demonstrated that pinned boundary conditions alter the global contractile response of the gel, it remains unclear whether stress relaxation occurs uniformly or through spatially heterogeneous deformation pathways. To address this, we analyzed the full strain tensor field and extracted the spatiotemporally resolved principal strain directions during contraction (Fig.~\ref{fig_results_director}a, white directors).

To quantify local variations during contraction, we introduce a metric that quantifies the spatially resolved strain direction field and compare free and pinned contraction cases. Specifically, we compute the angle 
$2\theta_p = \arctan\left(2\frac{\varepsilon_{xy}}{\varepsilon_{xx} - \varepsilon_{yy}}\right),$ 
where $\theta_p\equiv \theta_p(\mathbf{X},t)$ represents the orientation of the local principal strain axis at a point $\mathbf{X}$. As the angle  $2\theta_p$ is periodic in $\pi$, we use it to quantify the strain field director with nematic symmetry. 

To assess alignment of the overall sample, we evaluate $\cos(2\theta)$, where $\theta = 2\theta_p - \langle 2\theta_p \rangle_\mathbf{X}$ and $\langle \, \rangle_\mathbf{X}$ denotes spatial averaging and take values between $[-1,1]$. This metric measures the local deviation of strain orientations from the global contraction axis and removes head–tail ambiguity and measures alignment relative to a fixed reference axis.

For each frame, the spatially averaged orientational projection was computed as
$C(t) = \left\langle \cos(2\theta(\mathbf{X},t)) \right\rangle_{\mathbf{X}},$
where the average is taken over all interrogation windows within the region of interest. The temporal evolution of $C(t)$ was analyzed for each sample, and the resulting values across frames were summarized using box plots to compare fluctuations between control and pinned conditions across identified regions in strain evolution( Fig. ~\ref{fig_results_director}b). Values close to unity (1 or -1) correspond to strongly aligned deformation, whereas lower values closer to zero indicate larger local deviations in strain principal directions, reflecting increased spatial heterogeneity and non-affine strain evolution.

To quantify nematic ordering independent of the reference axis, we additionally computed the complex nematic order parameter
$Q(t) = \left\langle e^{i2\theta(\mathbf{X},t)} \right\rangle_{\mathbf{X}},
$
with magnitude
$
|Q(t)| =
\sqrt{
\left\langle \cos(2\theta) \right\rangle^2 +
\left\langle \sin(2\theta) \right\rangle^2
}$
The magnitude $|Q|$ measures the degree of orientational coherence independent of the dominant alignment direction and varies as $[0,1]$.
It quantifies angular concentration, taking the value $|Q|=1$
 when all angles are identical, $|Q|=0$
 when angles are uniformly distributed around the circle, and 
$0<|Q|<1$
 when the data exhibit partial clustering.

Figure~\ref{fig_results_director}b shows the distributions of $\langle \cos 2\theta \rangle$ during the lag, contraction, and stationary phases defined previously.The time traces of $\langle \cos(2\theta) \rangle$ fluctuated around zero for both control and pinned networks, indicating the absence of a persistent global alignment axis during contraction. However, the magnitude of these fluctuations differed substantially between conditions.
During the lag phase, both free and pinned systems exhibit relatively narrow distributions centered near zero, indicating minimal large-scale orientational organization prior to substantial contractile activity. Correspondingly, the box plots showed a smaller interquartile range and reduced spread, indicating weak fluctuations in orientational organization over time.
In the contraction phase, however, the behavior of the two systems diverges significantly.
Freely contracting samples maintain comparatively narrow distributions, indicating that most regions of the gel deform coherently along a common contraction direction.
In contrast, pinned samples exhibit substantially broader distributions with larger spread, an increased interquartile range, and pronounced outliers, indicating stronger local deviations from the mean strain orientation.
This qualitative broadening is especially evident in the larger variance of the pinned distributions during active contraction.
To quantify this observation, we performed Levene’s test comparing the variances of the free and pinned distributions during each contraction regime. During the contraction phase, the variances are statistically significantly different (p=0.0001), with pinned samples exhibiting substantially larger variance than freely contracting samples.
In contrast, no statistically significant difference is observed during the stationary phase, suggesting that strain heterogeneity emerges primarily during active stress buildup and mechanical relaxation rather than after contraction has saturated.

The observation that $\langle \cos(2\theta) \rangle \approx 0$ in both control and pinned systems indicates that neither condition exhibits a persistent global contraction axis. However, the fluctuations captured in the time traces and box plots reveal fundamentally different underlying dynamics.

Analysis of the nematic order magnitude $|Q|$ revealed that pinned networks exhibit stronger transient orientational coherence compared to controls. While the cosine projection remained centered near zero, intermittent increases in $|Q|$ indicate the emergence of spatially coherent anisotropic domains. This behavior is consistent with trends observed in $\langle \cos(2\theta) \rangle $, but importantly, the two conditions now show a statistically significant separation in the mean during the contraction phase. To quantify these observations, we performed Levene’s test to compare variances and a two-sample t-test to compare means between free and pinned distributions across each contraction regime. During the contraction phase, both the variances $(p = 1\times10^{-5})$ and means $(p = 1\times10^{-5})$ are significantly different, with pinned samples exhibiting substantially higher variance and mean values than controls. A consistent trend is also observed in the stationary phase, mirroring the behavior seen in $\langle \cos(2\theta) \rangle$ where we do not see any any significant differences in means or variances.

In control networks, the relatively small fluctuations suggest efficient relaxation of internally generated active stresses through filament rearrangement and network remodeling. This prevents the formation of long-lived anisotropic structures, resulting in weak temporal variability in orientational order.

In contrast, pinned networks exhibit larger fluctuations, indicating enhanced intermittency in orientational organization. The increased variability in strain directions observed in pinned samples can be understood as a consequence of mechanical frustration introduced by the boundary constraints. Mechanical confinement restricts large-scale stress relaxation and promotes accumulation of active stresses in localized regions. This leads to transient formation and dissolution of aligned domains, producing strong frame-to-frame variability.
Freely contracting gels interact minimally with the surrounding boundaries and therefore dissipate myosin-generated stresses through global shortening and cooperative network remodeling\cite{linsmeier_disordered_2016}.
In this case, motor activity continuously drives smooth deformation throughout the gel, resulting in relatively coherent strain alignment across the sample.
Pinned gels, however, are mechanically constrained by adhesion to opposing boundaries, preventing uniform contraction of the network. 
As myosin motors generate active tension, elastic stress accumulates internally within the gel.
Once the stored tension exceeds the local adhesion strength at weak points along the gel-wall interface, stress relaxation can occur through localized peeling or detachment events.
Simultaneously, different regions of the network relax unevenly, with some regions contracting strongly while others remain mechanically constrained.
These competing relaxation pathways generate spatially heterogeneous deformation patterns and promote non-affine strain evolution throughout the sample.
The broader strain-direction distributions observed in pinned samples therefore reflect the emergence of localized stress concentration, heterogeneous force transmission, and mechanically distinct relaxation domains within the active gel.

Importantly, a mean value near zero does not imply a fully isotropic state, as cancellation can arise from spatial heterogeneity or temporal switching of alignment direction. Thus, the variance and spread of $\langle \cos(2\theta) \rangle$ provide more meaningful information than the mean alone.

The nematic order parameter $Q$ resolves this ambiguity by capturing orientational coherence independent of direction. The increased fluctuations in $|Q|$ for pinned systems suggest that confinement stabilizes transient anisotropic structures and enhances mesoscale heterogeneity in stress transmission, consistent with intermittent processes such as rupture, filament sliding, and boundary untethering.

\subsection*{Pinned boundaries induce intermittent dynamics}

The non-uniform strain fields observed in the previous section suggest that stress relaxation in the actomyosin network does not occur homogeneously across the material.
Instead, contraction proceeds through spatially localized rearrangements and temporally evolving deformation pathways.
To further investigate the temporal persistence and evolution of these contractile states, we computed the autocorrelation of the evolving network configurations for both freely contracting and pinned systems.
This analysis enables quantification of how rapidly the network loses memory of its prior state and provides insight into the underlying relaxation dynamics governing active stress dissipation.

\begin{figure}[H]
	\centering
	\includegraphics[width=0.95\textwidth]{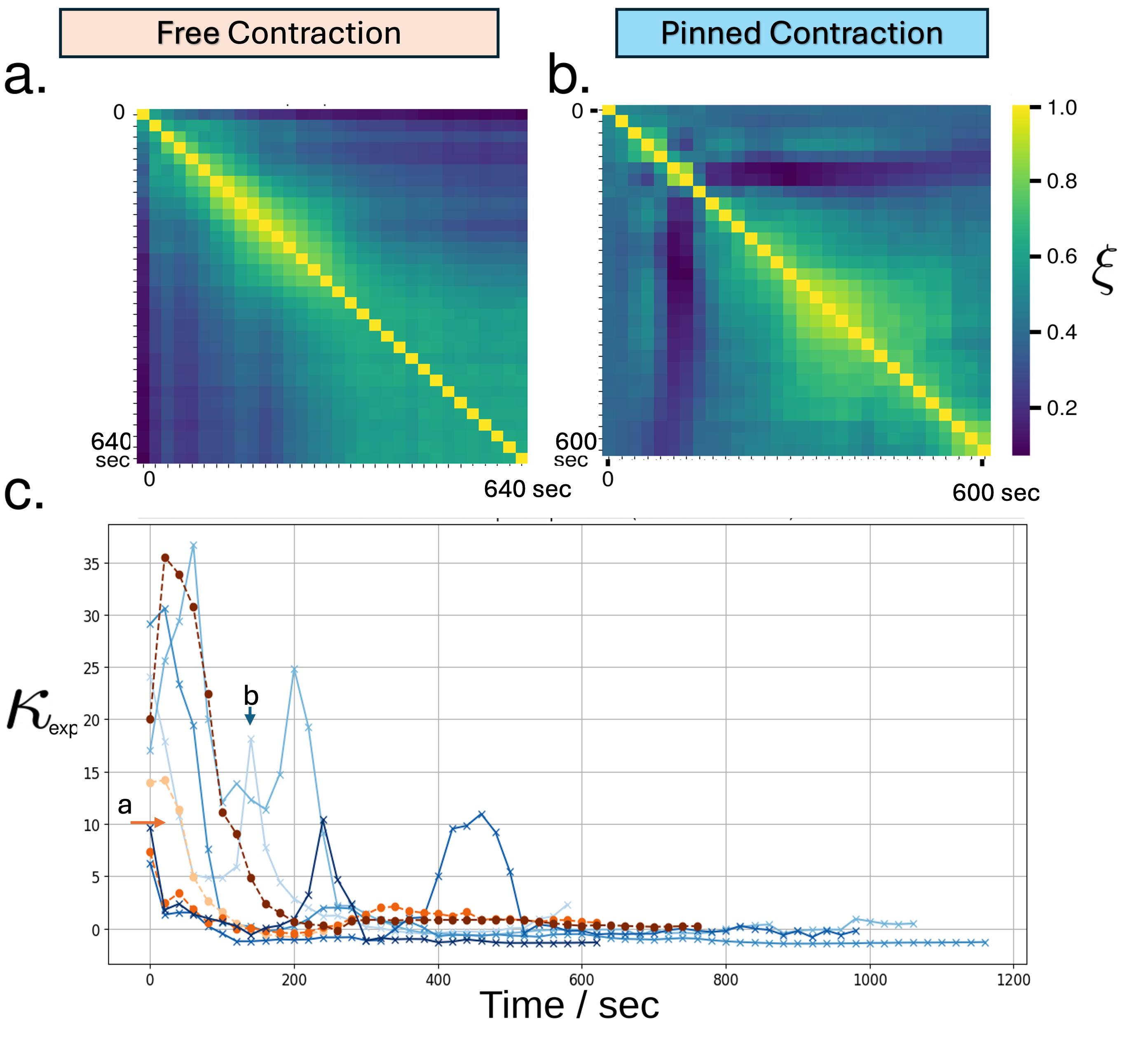}
	\caption{Pinned contraction results in intermittent dynamics. (a) Autocorrelation matrix of a Hencky strain field with color representing the correlation coefficient $\xi$ for a representative free contraction (control) experiment. (b) Autocorrelation matrix with color representing the correlation coefficient $\xi$ for a representative pinned contraction experiment. (c) Kurtosis, $\kappa_{exp}$, vs. time shown for all contraction experiments. The autocorrelation matrix shown in panel a and b corresponds to the arrowed experiments in panel c. }
	\label{fig_results_intermittent}
\end{figure}

For the freely contracting network (Fig.~\ref{fig_results_intermittent}a), the autocorrelation matrix was calculated by correlating network configurations across all pairs of time points during contraction.
The resulting matrix exhibits a smooth and gradual decay in correlation away from the diagonal, indicating continuous temporal evolution of the network structure.
Configurations separated by short time intervals remain strongly correlated, while correlations progressively weaken at longer separations.
Importantly, the matrix remains relatively continuous and homogeneous without abrupt discontinuities or strongly localized decorrelation bands.
This behavior suggests that freely contracting networks undergo progressive remodeling through distributed deformation and rearrangement mechanisms.
Active stresses generated by myosin motors are continuously dissipated through filament sliding, network reorganization, and collective contractile motion, resulting in a gradual loss of configurational memory.
The absence of sharp decorrelation events indicates that relaxation proceeds relatively smoothly, consistent with the continuously evolving strain fields observed earlier.
Together, these results support a picture in which free contraction enables large-scale cooperative rearrangement and homogeneous redistribution of active stress throughout the network.

In contrast, the autocorrelation matrix for the pinned contraction experiment (Fig.~\ref{fig_results_intermittent}b) exhibits markedly heterogeneous behavior.
While short-time correlations remain high near the diagonal, the matrix contains pronounced low-correlation bands and intermittent regions of abrupt decorrelation.
These features interrupt the otherwise gradual decay pattern and indicate the presence of sudden configurational changes during contraction.
Additionally, several regions remain correlated over comparatively long timescales, suggesting the coexistence of persistent structural domains alongside rapidly reorganizing regions.
The emergence of these heterogeneous correlation patterns indicates that pinning fundamentally alters the relaxation pathways available to the active network.
Mechanical constraints imposed by the pinned boundaries inhibit uniform contraction and prevent stresses from being redistributed globally.
As a result, elastic stress accumulates locally within the network until it is released through discrete rearrangement events such as filament slipping, rupture-like reorganization, or localized yielding.
The intermittent low-correlation bands therefore represent temporally localized stress-release events that abruptly change the overall network configuration.
Simultaneously, the persistence of long-lived correlated regions suggests that portions of the network become mechanically stabilized or trapped in metastable states due to the imposed constraints.

To quantify the intermittency of these dynamics, we calculated the kurtosis of the autocorrelation decay as a function of lag time (Fig.~\ref{fig_results_intermittent}c).
The red curves correspond to the freely contracting control network, while the blue curves correspond to the pinned contraction experiments.
The labeled markers “a” and “b” indicate the specific lag times associated with the autocorrelation matrices shown in Fig.~\ref{fig_results_intermittent}a and Fig.~\ref{fig_results_intermittent}b, respectively.
The freely contracting network exhibits relatively low kurtosis values that decay roughly monotonically over lag time, indicating that the relaxation dynamics remain comparatively uniform and distributed.
In contrast, the pinned system exhibits an initial decay followed by intermittent increases in kurtosis values, including sharp peaks and bursts at multiple timescales.
Elevated kurtosis reflects the presence of strong decorrelation events.
These peaks therefore indicate intermittent restructuring processes in which accumulated mechanical stress is suddenly released through localized rearrangements and depinning.
The non-monotonic kurtosis evolution observed in the pinned samples demonstrates that boundary constraints enhance not just spatial but also dynamical heterogeneity.
Rather than evolving through smooth continuous remodeling, the constrained network undergoes episodic restructuring events separated by periods of relative mechanical stability.
Such intermittency is characteristic of mechanically frustrated nonequilibrium systems and resembles avalanche-like relaxation or stick–slip yielding observed in other driven soft materials\cite{alvaradoUncoveringDynamicPrecursors2019}.
Taken together, the autocorrelation and kurtosis analyses demonstrate that boundary constraints strongly regulate stress relaxation pathways in active actomyosin gels.
Free contraction enables smooth and cooperative remodeling that continuously dissipates active stresses; whereas pinning localizes deformation, promotes stress accumulation, and drives intermittent restructuring events.
When interpreted alongside the previously observed non-affine strain fields, these results indicate that mechanical pinning transforms the network from a continuously remodeling active material into a dynamically heterogeneous system characterized by metastability, localized yielding, and temporally intermittent stress relaxation.

\subsection*{Built-up stresses are released via multiple relaxation pathways}

\begin{figure}[H]
	\centering
	\includegraphics[width=0.95\textwidth]{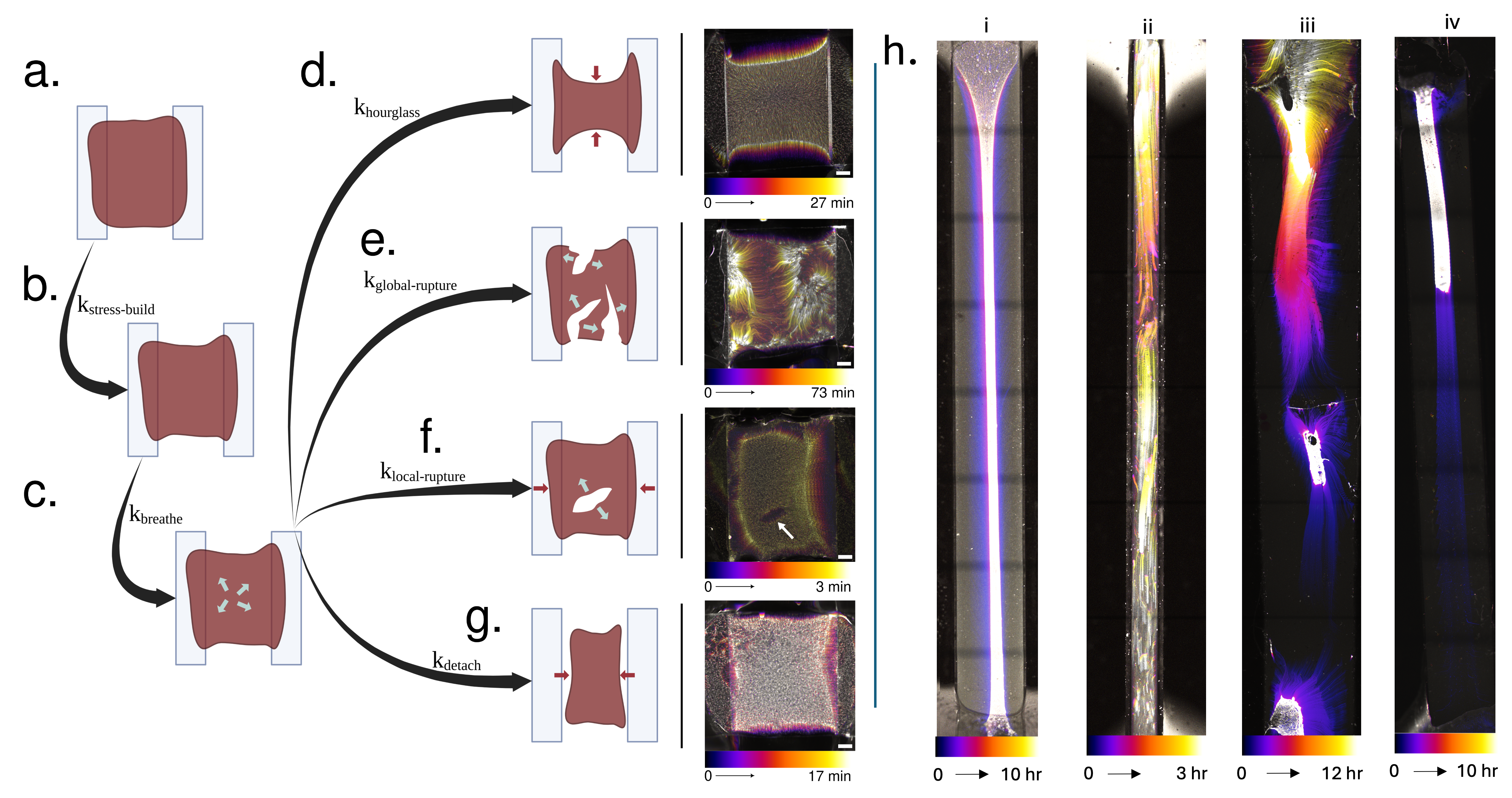}
	\caption{Experimental visualizations of pinned-contraction experiments, illustrating the distinct stress-relaxation pathways observed in this study. (a) A pinned actomyosin gel subject to transverse anchoring. (b) The first stage in contraction, where stress builds up within the gel. (c) The second stage in contraction, where the gel exhibits of internal motion. In the third stage of contraction, the gel exhibits various relaxation pathways including (d) hourglass (N=1), (e) global rupture (N=2), (f) local rupture (N=1), and (g) detachment from the boundary (N=5).  The local rupture time-color overlay depicts the moments after the gel breaks free and contains evidence for rupture (white arrow) within the internal meshwork of the gel. (h) Similar relaxation behaviour is seen when we have gels with different concentration and geometry. (i) (\qty{5.22}{\micro\molar} Actin, \qty{0.58}{\micro\molar} Tubulin, \qty{5.0}{\micro\molar} Myosin), (ii) (\qty{5.22}{\micro\molar} Actin, \qty{0.58}{\micro\molar} Tubulin, \qty{5.0}{\micro\molar} Myosin), (iii) (\qty{4.06}{\micro\molar} Actin, \qty{1.74}{\micro\molar} Tubulin, \qty{8.0}{\micro\molar} Myosin), and (iv) (\qty{5.8}{\micro\molar} Actin, \qty{2.5}{\micro\molar} Myosin) show hourglass (constriction), global rupture, local rupture, and detachment, respectively. Panels d--h are accompanied by time-color overlays of temporal epifluorescence micrograph image data.For all time-color overlay micrographs, the scale bar is \qty{500}{\micro\meter}.}
	\label{fig_results_relaxations}
\end{figure}

To investigate how boundary constraints influence the mechanical evolution of active gels, we next examine the distinct stress-relaxation pathways observed in pinned-contraction experiments. Here, we observe all $N=9$ experiments from our study. 
Figure~\ref{fig_results_relaxations} summarizes the range of dynamical responses that we observed.

The pinned-contraction experiments were visualized using temporal epifluorescence microscopy combined with time-color overlay imaging to capture the evolution of gel morphology during contraction.
Across the dataset, we identify three general stages of evolution: an initial stress build-up phase (Fig.~\ref{fig_results_relaxations}a,b), an intermediate regime characterized by intermittent internal motion or “breathing”-like dynamics (Fig.~\ref{fig_results_relaxations}c), and a final stress-relaxation stage in which the gel releases accumulated stresses through one of several distinct mechanical pathways (Fig.~\ref{fig_results_relaxations}d–g).

One frequently observed pathway is symmetric hourglass formation (Fig.~\ref{fig_results_relaxations}d), in which the gel progressively narrows at its center while remaining attached to the boundaries. The deformation remains remarkably stable, with some samples maintaining the hourglass morphology for up to \qty{27}{min} before rupturing at one of the anchoring points. Similar necking-like behavior was also reproduced in simulations, where its stability was found to depend strongly on the adhesion strength at the boundaries. Stronger adhesion promotes sustained stress accumulation by preventing premature detachment, thereby delaying rupture and stabilizing the constricted morphology. The long-lived nature of the hourglass state suggests that pinned boundaries enable the network to store substantial elastic stress prior to mechanical failure.

A second relaxation pathway involves localized internal rupture within the gel bulk (Fig.~\ref{fig_results_relaxations}e). In this case, a single rupture nucleates inside the network rather than at the boundaries, indicating that stress concentration can develop internally under constrained contraction. Following rupture formation, the surrounding network recoils perpendicular to the rupture axis, consistent with rapid local stress relaxation. Unlike cracks in brittle solids, however, the rupture does not continue to propagate longitudinally after nucleation. Instead, its length remains relatively fixed following formation. This behavior suggests that stress redistribution and subsequent boundary detachment relieve the driving force required for continued crack propagation, distinguishing these active gels from conventional brittle fracture systems.

We additionally observe global rupture events (Fig.~\ref{fig_results_relaxations}f), where the gel fragments into multiple disconnected clusters through several simultaneous fractures. These catastrophic failure modes resemble the dynamics previously reported in actomyosin systems near critical connectivity thresholds\cite{alvaradoMolecularMotorsRobustly2013}. In contrast to localized rupture, global rupture reflects a system-wide loss of mechanical integrity and indicates that stresses have become sufficiently large to destabilize the network collectively.

The most commonly observed relaxation pathway corresponds to boundary detachment (Fig.~\ref{fig_results_relaxations}g), in which the gel remains pinned for a finite period before abruptly peeling away from one or both anchoring surfaces. This pathway was observed in five independent samples and formed the basis of the earlier energetic characterization. The delayed detachment behavior further supports the idea that transverse adhesion suppresses immediate contraction and instead permits progressive accumulation of internal active stresses before sudden release.

Importantly, the rupture times associated with the hourglass and global rupture pathways are substantially longer than the contraction timescales observed in freely contracting control gels. This extended timescale demonstrates that mechanical confinement and adhesion fundamentally alter the relaxation dynamics by enabling the network to store elastic energy over prolonged durations prior to failure. Together, these observations reveal that pinned actomyosin gels do not relax through a single deterministic mechanism, but instead access multiple competing stress-relaxation pathways governed by the interplay between active contractility, network remodeling, and boundary adhesion.

% ...

Finally, to assess whether the observed stress-relaxation pathways are specific to quasi-2D actomyosin gels or represent more general features of confined active matter, we extend our experiments to actomyosin–microtubule composite gels prepared at varying concentrations and confined within long cylindrical glass capillaries (Methods). This geometry introduces a distinct confinement condition while preserving the ability of the network to undergo active contraction and stress redistribution, allowing us to probe the robustness of the relaxation mechanisms identified above.

Figure~\ref{fig_results_relaxations}h shows representative time-color overlay visualizations of contraction dynamics in these composite systems. Despite differences in composition and geometry, we again observe the same characteristic relaxation pathways identified in the quasi-2D actomyosin gels, including hourglass-like constriction, global rupture, localized rupture, and boundary detachment (Fig.~\ref{fig_results_relaxations}h,i–iv). The preservation of these distinct dynamical outcomes across systems suggests that the underlying relaxation mechanisms are not sensitive to specific biochemical composition or confinement geometry, but instead emerge from generic features of active stress generation coupled to mechanical constraints.

The agreement between these composite-gel dynamics and those reported in purely actomyosin networks demonstrates that the identified stress-relaxation pathways constitute robust emergent behaviors of confined active gels. In particular, the recurrence of similar morphological evolution across varying concentrations indicates that these pathways are governed primarily by the interplay between active contractility, network connectivity, and boundary-mediated stress accumulation rather than fine biochemical details.

Preliminary observations from the localized rupture pathway further reveal signatures of propagating relaxation fronts following failure events (Supplemental Fig.~4). These dynamics suggest that stress release is not purely local but can propagate through the network in a wave-like manner during rupture, providing additional evidence for collective mechanical coupling in active-gel failure processes.


\section*{Discussion}

Pinned contraction events resemble the tension-bearing supracellular actomyosin network of tissues and developing embryos undergoing morphogenesis. Recent advances in colloidal nanocrystal gels and dynamic covalent hydrogels have demonstrated that structural evolution and bond kinetics strongly regulate stress transmission and relaxation \cite{crowell_leveraging_2026,ofosu_universal_2026,kang_colloidal_2025}, suggesting analogous mechanisms may underlie the heterogeneous strain localization and stress-relaxation pathways observed in ATP-driven actomyosin networks.
The mechanical constraints that transverse adhesion places on contractile gels results in a variety of mechanical phenomena not observed in free contraction.
We have shown that pinned contraction events exhibit stress-build up (cf. Fig.~\ref{fig_results_strain},\ref{fig_results_energy}).
These these stresses coincide with spatial (cf. Fig.~\ref{fig_results_director}) and temporal (cf. Fig.~\ref{fig_results_intermittent}) inhomogeneities in the strain field, which we quantified for samples that detached from the boundary.
Meanwhile, we observed instances of additional stress-relaxation behaviors, such as hourglass deformations and rupture (cf. Fig.~\ref{fig_results_relaxations}).
These properties place tension-bearing, pinned active gels in a unique category of active matter with properties that warrant additional study.

\paragraph*{Properties of tension-bearing active gels.} The intermittent dynamics we report (cf. Fig.~\ref{fig_results_intermittent} and Fig.~\ref{fig_results_strain}) manifest as stick-slip-like motion, characterized by periods of slow buildup of stress followed by sudden stress-release events. These dynamics are reminiscent of failure precursors observed in other soft and biological materials under load,\cite{aimeMicroscopicDynamicsFailure2018, alvaradoUncoveringDynamicPrecursors2019} and are indicative of the system approaching a mechanical instability or depinning transition.

Such intermittent behavior is a signature of systems with internal friction and structural heterogeneity, where local rearrangements or ruptures can propagate and trigger global relaxation events. In our system, these dynamics likely serve as a mechanism for stress dissipation, especially in the presence of boundary adhesion that resists contraction. Similar stick-slip-like behaviors have been reported in epithelial tissues,\cite{serra-picamal_mechanical_2012} active nematics,\cite{duclos_spontaneous_2018} and in vitro actomyosin networks under load.\cite{murrell_f-actin_2012}

In addition to intermittency, we observe pronounced strain non-uniformity across the gel, suggesting spatially heterogeneous stress distributions. These non-uniform strain fields are a hallmark of tension-bearing gels and are absent in freely contracting systems where stress is more uniformly distributed. Strain gradients are closely tied to stress gradients, which in biological systems are known to drive cortical flows and contribute to processes such as cell polarization and tissue morphogenesis\cite{mayerAnisotropiesCorticalTension2010}. The presence of these gradients in our system further supports the relevance of our reconstituted gels as a model for understanding mechanical behavior in active biological materials.

\paragraph*{Pinned boundaries and caged ATP as tools to mimic cellular stress regulation} In our reconstituted actomyosin gels, the use of pinned boundaries with adhesion mimics the mechanical constraints experienced by cells within tissues. In vivo, cells adhere to their surroundings via integrin-based focal adhesions or cadherin-mediated cell-cell junctions, which serve as mechanical anchors that resist contractile forces and transmit tension across tissues \cite{discherTissueCellsFeel2005a,Lecuit2007}. Similarly, in our system, boundary adhesion provides a mechanical constraint that resists contraction, enabling the buildup of internal stress and the emergence of strain gradients. This setup allows us to model how cells and tissues manage mechanical stress through anchoring points.


In addition, the use of caged ATP provides a means to temporally control the onset of contractility. Upon photorelease, ATP activates myosin motors, initiating contractile activity in a synchronized manner across the gel. This approach enables precise probing of the dynamics of stress buildup and relaxation, and mimics the spatiotemporal activation of contractility observed in vivo, such as during morphogenetic events like apical constriction\cite{martinPulsedContractionsActin2009}.


\paragraph*{Differences between transverse anchoring and lateral anchoring} In previous work with actomyosin gels, buildup of stress and structural changes during contraction have been observed in the context of boundary coupling,\cite{bendixQuantitativeAnalysisContractility2008a,jia3DPrintedProteinbased2022} highlighting the interplay between internally generated stress and boundary adhesion. In this context, we distinguish between two relevant types of boundary conditions: lateral anchoring and transverse anchoring.

Lateral anchoring is most relevant when modeling actomyosin behavior in individual cells. It can be conceptualized as frictional coupling between the actin cortex and the plasma membrane, which attenuates contraction length scales\cite{murrellFactinBucklingCoordinates2012,alvaradoChapterReconstitutingCytoskeletal2015}. This friction can be relaxed either through cortical rupture\cite{murrellFactinBucklingCoordinates2012} or via cortical flows driven by tension gradients\cite{mayerAnisotropiesCorticalTension2010}. Such flows are essential for establishing cell polarity and driving morphogenetic processes during development. When the cortex is supported by a solid substrate rather than a fluid-like membrane, stress relaxation often occurs through internal ruptures that fragment the gel\cite{alvaradoChapterReconstitutingCytoskeletal2015}.

Transverse anchoring, by contrast, refers to a configuration where a contractile gel is anchored at two opposing ends and allowed to relax freely in between. This geometry is particularly relevant for modeling supracellular actomyosin networks in tissues, where contractile cables span multiple cells and are anchored at junctions\cite{roperSupracellularActomyosinAssemblies2013}. In our current study, we implement transverse anchoring by pinning the gel at its ends, mimicking cell-cell adhesion in tissues. This setup allows us to observe stress buildup, strain non-uniformity, and intermittent dynamics - hallmarks of tension-bearing active materials.

Furthermore, by combining transverse anchoring with caged ATP activation, we mimic both the mechanical loading and biochemical triggering observed in vivo. This approach enables us to probe how stress relaxation unfolds in a controlled environment, revealing dynamic behaviors such as stick-slip-like motion and rupture events. These features are reminiscent of tissue-scale contractions during processes like Drosophila gastrulation \cite{decker_tension_2025}, where a rectangular domain of cells activates Rho1 to drive coordinated apical constriction over a timescale of 10–15 minutes\cite{richRho1ActivationRecapitulates2020,sweetonGastrulationDrosophilaFormation1991}.

Our findings suggest that transverse anchoring, when combined with spatiotemporal control of contractility, provides a powerful framework for studying the mechanics of active gels and their relevance to tissue morphogenesis.

\paragraph*{Timescales of relaxation} The relaxation of stress in actomyosin networks is governed by a combination of biochemical and mechanical processes. In vivo, actin filament turnover, driven by polymerization, depolymerization, and severing, contributes to rapid remodeling of the cytoskeleton on timescales of minutes\cite{fritzscheAnalysisTurnoverDynamics2013,wigbers_flow_2020}. Severing proteins such as cofilin and gelsolin can rapidly disassemble filaments, enabling fast redistribution of stress. These biochemical processes are complemented by mechanical relaxation mechanisms, including rupture, depinning, and reorganization of the network.
Interestingly, the timescales of intermittent dynamics observed in our system are also on the order of seconds to minutes, consistent with prior studies of active gels and living cells\cite{silvaTimeresolvedMicrorheologyActively2014,humphrey_active_2002}. This suggests that mechanical and biochemical relaxation processes may operate on overlapping timescales, potentially interacting to regulate stress dissipation.

In more stable, tension-bearing structures such as stress fibers, actin filaments are known to persist longer and resist depolymerization\cite{tojkanderActinStressFibers2012}. In these contexts, repair mechanisms become critical. For example, the LIM-domain protein Zyxin has been shown to localize to sites of mechanical damage in actin filaments, promoting their repair and reinforcing the network\cite{smithZyxinMediatedMechanismActin2010,hoffman_dynamic_2011,phuaForceactivatedZyxinAssemblies2025}. Moreover, stressed actin filaments have been shown to be protected from severing, suggesting a feedback mechanism where mechanical load stabilizes the cytoskeleton\cite{hayakawa_actin_2011}.

In our system, the presence of phalloidin stabilizes actin filaments by preventing depolymerization, effectively suppressing filament turnover. As a result, the dominant relaxation pathways are likely limited to mechanical rupture and rearrangement, rather than biochemical turnover. This constraint may bias the observed dynamics toward larger-scale rupture events, as opposed to more distributed, filament-level remodeling.
Furthermore, factors such as filament alignment, crosslinking heterogeneity, and spontaneous defect formation all contribute to stress relaxation\cite{murrellFactinBucklingCoordinates2012,alvaradoChapterReconstitutingCytoskeletal2015}.

The variety of stress relaxation pathways we observed (cf. Fig.~\ref{fig_results_relaxations}) resembles those previously in confined liquid-crystal tactoid systems, suggesting that active soft materials under geometric or adhesive constraints can access multiple metastable relaxation modes \cite{garleaColloidalLiquidCrystals2019a}.
We anticipate that multiple dynamical processes within the gel control the ocurrence of relaxation pathways, including active stress generation, detachment rate, stress propagation rate, and viscoelastic response.
Furthermore, recent studies in chemical systems demonstrate that control over reaction rates controls the resulting non-equilibrium chemical states.

One direction for future work would be to compare phalloidin-stabilized gels with dynamic gels that allow filament turnover, to assess how biochemical remodeling influences stress relaxation. It is possible that in more physiological conditions, smaller-scale ruptures or filament severing events allow for more distributed and less catastrophic stress dissipation. Such mechanisms are thought to be critical in tissues, where controlled relaxation prevents mechanical failure and enables robust morphogenesis \cite{wyattQuestionTimeTissue2016,khalilgharibiStressRelaxationEpithelial2019,gupta_adaptive_2015}.

\paragraph*{Comparing experimental and simulation relaxation pathways.}
Within our simulations, we were only able to observe a subset of the relaxation pathways observed in experiment. We observed similar necking behavior, and - in cases where the boundary adhesion strength was weak enough - we observed depinning events that were qualitatively, and quantitatively similar to our experimental findings. (cf. Figures \ref{fig_results_strain},\ref{fig_results_energy},\ref{fig_results_intermittent}.) For the internal rupture observations, future work could explore incorporating defects into simulated network interiors to investigate whether this leads to this relaxation pathway.

\paragraph*{Limitations of the linear hydrodynamic model.} In our study, we use an effective activity parameter as a proxy to estimate the evolving mechanical state of the gel, employing a simplified linear hydrodynamic stress model. This approach captures essential features of stress relaxation by incorporating linear contributions from elastic, viscous, and decaying active stress terms. While this model provides a tractable framework for interpreting experimental data and identifying key trends, it necessarily omits many of the nonlinear and spatially heterogeneous features intrinsic to active cytoskeletal materials.

The stress model assumes that the total stress $\sigma$ is governed by a linear superposition of elastic strain $\epsilon$, viscous dissipation $\frac{d\epsilon}{dt}$, and an exponentially decaying active stress term $\zeta e^{-t} \epsilon$. This formulation allows us to relate stress relaxation to changes in an effective activity parameter without explicitly modeling the underlying molecular interactions. Such simplifications are common in active matter modeling,\cite{prost_active_2015,julicher_hydrodynamic_2018} and are useful for capturing macroscopic behavior. However, they come with limitations.

Active stress in cytoskeletal gels arises from a complex interplay of biochemical and mechanical processes, including ATP hydrolysis, myosin motor activity, filament alignment, crosslinking dynamics, and spatial heterogeneity in motor distribution\cite{salbreuxActinCortexMechanics2012,kruseGenericTheoryActive2005}. Our model approximates these contributions using a single decaying activity term, which captures the overall relaxation profile but neglects spatial and temporal variations in activity. For instance, gradients in ATP concentration,\cite{bement_activatorinhibitor_2015} local myosin clustering, and filament buckling\cite{murrell_f-actin_2012} can all lead to localized contractility and non-uniform stress propagation, which are not resolved in our current framework.

Despite these simplifications, our model successfully captures the transition from active to passive relaxation regimes. Non-dimensionalizing the stress equation yields:
\[
\sigma = \epsilon + \Pi \frac{d\epsilon}{dt} + \zeta' e^{-t} \epsilon,
\]
where $\sigma = \sigma/G$, $t = \beta t$, $\Pi = \eta \beta / G$, and $\zeta' = \zeta/G$. This leads to a characteristic relaxation timescale:
\[
t_{\text{relax}} = \frac{\Pi}{1 - \zeta'}.
\]
At short timescales, relaxation is dominated by the decaying activity parameter, while at longer timescales, the system behaves like a passive viscoelastic gel, effectively reducing to a Maxwell model. This crossover is consistent with observations in both reconstituted and cellular systems, where active stress dominates early dynamics, followed by passive dissipation\cite{mizunoNonequilibriumMechanicsActive2007,humphrey_active_2002}.

Future work could extend this model by incorporating spatially resolved activity fields, nonlinear elasticity, and feedback between stress and biochemical signaling. Such extensions would allow for a more complete description of the rich spatiotemporal dynamics observed in active gels and tissues.

\paragraph*{Mesoscopic density heterogeneities} Contractile active gels consist of a dense ensemble of actin filaments, bundles, and molecular motors that continuously reorganize under internally generated stresses. As a result, these systems naturally develop mesoscopic density heterogeneities on length scales of approximately \qtyrange{1}{10}{\micro\meter}. Such heterogeneities emerge from the interplay between active contractility, filament transport, bundling, and local stress redistribution, and therefore provide insight into the underlying mechanical state of the network.

In freely contracting systems, active stresses can be dissipated through global shortening and large-scale network rearrangement, allowing density fluctuations to relax continuously during contraction. In contrast, mechanically constrained or pinned networks are expected to accumulate internal tension, potentially enhancing spatial density variations through localized compaction, stress concentration, and heterogeneous filament transport. These processes may generate dynamically evolving dense and dilute regions within the gel, reflecting nonequilibrium restructuring pathways associated with stress buildup and release.

\section*{Conclusion}


We examine contractile actomyosin active gels under Dirichlet boundary conditions ($U=0$ at both edges), mimicking adhesion to rigid substrates. Pinned boundaries suppress global contraction and generate heterogeneous displacement and strain fields.
Hencky strain evolution shows slow contraction interrupted by propagating tensile and contractile zones with asymmetric, non‑affine deformations, indicating that structural defects modulate local force transmission. Boundary pinning localizes strain and delays contraction, while internal heterogeneity redirects stress and shapes relaxation pathways.
Pinned and freely contracting gels exhibit distinct relaxation behaviors, including detachment‑mediated relaxation in pinned samples. These differences appear in strain evolution and in activity‑dependent relaxation rates captured by our hydrodynamic model, which resolves changes in relaxation before and after detachment.
Intermittent dynamics, strain non‑uniformity, and rupture events further highlight the complex mechanical response of tension‑bearing actomyosin networks. These behaviors are relevant for understanding stress relaxation in cellular and tissue‑scale assemblies where active forces interact with mechanical constraints.
Systematic in vitro studies under varied boundary conditions will help elucidate principles of active stress regulation and inform the design of soft robotic systems that exploit active materials for adaptive mechanical responses.
\section*{Acknowledgments}
This research was primarily supported by the National Science Foundation through the Center for Dynamics and Control of Materials: an NSF MRSEC under Cooperative Agreement No. DMR-2308817. This research was supported in part by grant NSF PHY-2309135 and the Gordon and Betty Moore Foundation Grant No. 2919.02 to the Kavli Institute for Theoretical Physics (KITP). This research was supported in part by the National Science Foundation CAREER Grant No. DMR-2144380 and National Institutes of Health under award number R21HD112657. We acknowledge funding from the US National Science Foundation DMREF program through following grants: NSF DMR-2119663 (to RMRA), NSF DMR-2118497 (to MTV), NSF DMR-[Mo's  here] (to MD). 

\section*{Supplementary materials}
Methods

 Imaging Protocol
 Protein Preparation

Supplemental Figure 1 - Full assay chamber overview\\
Supplemental Figure 2 - Simplified assay chamber overview\\
Supplemental Figure 3 - Assay chamber construction\\
Additional results\\
Supplemental Figure 4 - Global vs Local strain\\
Supplemental Figure 5- Wave Speed\\
Supplemental Figure 6 - Strain Field snapshot in lag phase\\
Supplemental Figure 7 - Rise Time\\
Supplemental Figure 8 - Lag Time\\
References \textit{(List SI citations here)}

\section*{Author contributions}
A.M. is the main contributor to this work, leading the analysis, modeling, and writing of the manuscript. J.A. and J.C. conceptualized the study. J.C., H.L.,A.M. and K.W. conducted the experiments. H.L. and A.M. contributed to the analysis of experimental results. A.M. led the modeling efforts. J.C., A.M., and J.A. contributed to writing the manuscript. H.L. and K.W. assisted in writing and provided feedback on several sections. A.R. and M.D. contributed for simuations. A.S. and R.RA. provided their experimental data for comparison.

\bibliographystyle{unsrt}
\bibliography{biblio}

\clearpage

\end{document}


\let\oldthebibliography=\thebibliography
\let\oldendthebibliography=\endthebibliography
\renewenvironment{thebibliography}[1]{
    \oldthebibliography{#1}
    \setcounter{enumiv}{57}                        
}{\oldendthebibliography}
\title{Supplementary Material}

\maketitle


\section*{Methods}

\subsection*{Imaging protocol.}
The samples are imaged in a Zeiss Axio Observer 5 epifluoresence microscope with a broad spectrum X-Cite Xylis LED in twenty second intervals with two optical channels at 15\% intensity and 60\% intensity through a Zeiss 64 HE filter and 49 filter, with an exposure time of 5 milliseconds and 50 milliseconds, respectively. Each experiment is imaged for a total of 30 minutes. The excitation windows are between 300 nm to 395 nm and 570 nm to 600 nm, with emission windows between 410 nm to 460 nm, and 605 nm to 680 nm for the 49 filter and the 64 HE filter, respectively. The DAPI filter excites NADH, and the PLUM filter excites the AlexaFluor dye bonded to actin filaments.

Imaging is performed on a Zeiss Axio-observer inverted epifluorescence microscope with Teledyne Photometrics Prime BSI Scientific CMOS camera and Lumen Dynamics X-cite Xylis LED fluorescence microscope light source. We image using mPlum (actin) and DAPI (caged-ATP release) filters with a EC Plan-NEOFLUAR 2.5x NA 0.085 objective (Zeiss). mPlum intensity is set to $60$\% with $2.5ms$ exposure. mPlum images are acquired every $20s$. DAPI intensity is set to $100$\% with variable exposure from $500ms$ to $10s$. For high throughput experiments, a custom Python based control system is utilized based on the OAD platform from Zeiss. Analysis is performed using Python with standard libraries.

\subsection*{Protein preparation.}
Monomeric (G-) actin and myosin II were purified from rabbit psoas skeletal muscle [297]. G-actin was purified using a Superdex 200 column (GE Healthcare, Waukesha, WI, USA) and stored at -80°C in G-buffer (2 mM tris-hydrochloride, pH 8.0, 0.2 mM disodium adenosine triphosphate, 0.2 mM calcium chloride, 0.2 mM dithiothreitol). Myosin II was stored at −20°C in a high-salt storage buffer containing glycerol (25 mM monopotassium phosphate, pH 6.5, 600 mM potassium chloride, 10 mM ethylenediaminetetraacetic acid, 1 mM dithiothreitol, 50\% w/w glycerol). Creatine phosphate disodium and creatine kinase were purchased from Roche Diagnostics (Indianapolis, IN, USA), and all other chemicals were obtained from Sigma Aldrich (St. Louis, MO, USA). Magnesium adenosine triphosphate was prepared as a 100 mM stock solution using equimolar amounts of disodium adenosine triphosphate and magnesium chloride in 10 mM imidazole, pH 7.4. Myosin II was labeled with Alexa Fluor 488 NHS ester (Invitrogen, Paisley, UK), and actin was labeled with Alexa Fluor 594 carboxylic acid, succinimidyl ester [297]. Mouse fascin was purified by transforming T7 \textit{Escherichia coli} bacteria with a glutathione S-transferase (GST)-fascin pGEX vector (Gentry et al., 2012). The protein was expressed with a GST tag and purified by cleaving the tag while bound to a glutathione sepharose column (GE Healthcare). Fascin protein was snap-frozen in aliquots and stored at -80°C in 20 mM imidazole, pH 7.4, 150 mM potassium chloride, 1 mM dithiothreitol, and 10\% v/v glycerol.
100 uL of Myosin II Protein was dialyzed overnight in 100 mL of Myosin Dialysate (20 mM imidazole pH 7.4, 300 mM potassium chloride, 4 mM magnesium chloride, 1 mM dithiothreitol-tol), at 4°C with a stir rod rotating at about 180 rpm. 100 uL of the dialysate and 100 uL of the dialyzed myosin was saved in Eppendorf tubes and kept over ice in a 4°C fridge.
A chosen volume of actin and arp ⅔ are each dialyzed in a beaker of 100 mL GB\textsuperscript{-1000x }(5 mM tris-hydrochloride, pH 7.8, 0.2 mM disodium adenosine triphosphate, 0.1 mM calcium chloride, 5 mM dithiothreitol) overnight at 4°C with a stir rod rotating at about 180 rpm. Alliquotes of actin and arp 2/3 are appropriately distributed into Eppendorf tubes and flash frozen using liquid nitrogen to be stored at -80°C for future use.

\subsection*{Glass slide and coverslip preparation.}
To prevent any non-specific interactions and remove any impurities that may affect the results of the experiment, the coverslips and glass slides undergo a sonification and base piranha solution cleaning process. Both items undergo the same treatment, but are handled in different carousels to accommodate their respective sizes.
Glass coverslips (22mm x 40 mm, thickness \#1 ¼ 0.15 mm, Menzel Glaser) are loaded into a Teflon carousel, while glass slides are loaded into their own carousel. The carousels containing the coverslips and glass slides are each placed into clean beakers filled with Milli-Q water. They are sonicated using a bath sonicator for 5 minutes at room temperature. The carousels are drained and the coverslips and glass slides are blow-dried with a stream of clean nitrogen gas.
The coverslips and glass slides are reloaded into their respective carousels. Two beakers, each filled with five parts Milli-Q water, are heated on a hot plate to 80°C while being continuously stirred with a magnetic bar. While under a fume hood, one part ammonium hydroxide (30\% stock) and one part hydrogen peroxide (30\% stock) is added to each beaker. Upon mixing, the solutions spontaneously formed bubbles, indicating an active reaction. The amount of base piranha solution should suffice to completely submerge a carousel and its contents. The carousels were each set in a beaker and incubated in the piranha solution for 30 minutes maintaining 80°C. The beakers were loosely covered with aluminum foil to prevent splashing while allowing fumes and water vapor to escape.
The coverslips and glass slides were once again rinsed with milli-Q, dried with nitrogen gas, loaded into a clean carousel, and are each entirely immersed in a Milli-Q filled beaker to be sonicated for 5 minutes at room temperature. The carousels were drained and the coverslips and glass slides were dried with nitrogen gas. After reloading each carousel, they were filled with isopropanol until the contents were covered entirely. Both were tightly sealed with parafilm to prevent evaporation. Once needed, clean tweezers are used to remove either a glass slide or coverslip and are dried with nitrogen gas.
\begin{figure}
    \centering
    \includegraphics[width=0.75\linewidth]{Screenshot 2025-01-30 103118.png}
    \caption{ Visual representation of chamber design for pinned chamber contractions}
    \label{fig:enter-label}
\end{figure}
    A clean glass slide [1.2] was covered with a parafilm layer, and the design shown in Figure 1 was etched into the film using a Trotec Speedy 360 CO2 Laser. The outline was then traced with a clean scalpel. The top and bottom strips of 75 mm x 3 mm film were removed and disposed of, along with the thicker solid middle strips measuring 3.25 mm x 19 mm and 3.50 mm x 19 mm. The thinner outer solid strips, measuring 2 mm x 19 mm, were removed and stored in a sterile petri dish for later use. The 1 mm x 5 mm semi circles were removed and disposed of, while the outer four and inner two 3 mm x 19 mm chamber outlines on either side were removed and stored in a sterile petri dish for later use. The final outline is shown in Figure 1.3.2.
\begin{figure}
    \centering
    \includegraphics[width=0.75\linewidth]{Screenshot 2025-01-30 103156.png}
    \caption{ Visual representation of final chamber outline }
    \label{fig:enter-label}
\end{figure}
 Wings made of acrylic or 316L steel were added to each chamber, oriented in pairs such that the notched side walls were positioned over the non-elliptically etched side. A 3 mm x 19 mm chamber-outlined Parafilm strip, taken from the outer Parafilm cuts, was placed over both wings so that the semicircles aligned. A clean glass coverslip [1.2] was cut to the approximate dimensions of 12 mm x 5 mm and placed on top of the chamber, with the edges of the glass overlapping well with the Parafilm wing-Parafilm stack on either side. A thin 2 mm x 19 mm solid piece of stored Parafilm was then added to span the coverslip, ensuring it did not cover the elliptical notch. The glass slide carrying the chambers was set onto a hotplate heated to 120°C. Gentle pressure was applied to the chamber's perimeter to secure the Parafilm, wing, and coverslip layers, avoiding the semicircles to maintain the chamber's shape. The glass slide was removed from the heat once the layers were secure and the Parafilm appeared clear. The final slide was stored in a sterile labeled petri dish.
 \begin{figure}
     \centering
     \includegraphics[width=0.75\linewidth]{Screenshot 2025-01-30 103210.png}
     \caption{ Visual progression of chamber being built }
     \label{fig:enter-label}
 \end{figure}

\subsection*{Protein clarification.}
Prepared proteins from [1.1] involving labeled actin, dark actin, myosin, fascin, arp ⅔, profilin, and GST-VCA were spun in a centrifuge to remove aggregates. Frozen pre-prepared aliquots of labeled actin, dark actin, fascin, arp ⅔, profilin =, and GST-VCA were removed from a -80C freezer and thawed. All proteins were spun at 55,000 rpm, at 4C for 5 minutes using a . The supernatant was moved to Eppendorf tubes via pipette and kept on ice.

%
The protein solutions absorbances at different wavelengths are measured using a NanoDrop 2000. The table above shows the blanks and wavelengths used for each reagent. The measured absorbances and respective molar absorbance coefficients are used to calculate the concentration of protein in each solution using the Beer-Lambert law.
    
\begin{equation*}
    A =  \varepsilon c L
\end{equation*}
                                       \caption{Figure 1.5.2: Beer-Lambert Equation}

\subsection*{Chamber passivation.}
One chamber of the two created was chosen as the pinned chamber, while the other served as the control chamber. The both chambers were rinsed alternately between EtOH and Milli-Q using nitrogen gas in between each, until no air bubbles were seen and the final wash was Milli-Q. The chambers were then slightly overfilled with KOH (1.0 M) and set into a petri dish with a kimwipe oversaturated with Milli-Q for 10 minutes. Using nitrogen gas, the chamber is cleared from KOH and rinsed with Milli-Q. These steps rinsed out organic substrates that could cause non-specific binding. The chambers were then overfilled with PLL-PEG (0.2 mg/mL of Milli-Q) and incubated in a petri dish with a kimwipe oversaturated with Milli-Q for 45 minutes. Finally, the chambers were rinsed with Milli-Q and dried with nitrogen gas.

After Chemical attachment of PLL-PEG, a mask was used to isolate the winged sides of the pinned chamber walls and expose them to UV light (254 nm) for 5 minutes on both sides of the glass slide. This opens up reactive groups on the PLL-PEG surface, allowing for covalent linkages, electrostatic binding, or strong non-covalent interactions. 

After exposure, the pinned chamber was overfilled with GST-VCA (3 μM) for 15 minutes in a petri dish with a kimwipe oversaturated with Milli-Q. In addition to the standard contractile assay reagents, pinned experiments utilize Arp2/3 Complex, Profilin, and GST-VCA. These proteins work in concert to help ensure that actin nucleates at the boundary where the GST-VCA had adsorbed. Assays are formulated with [Arp2/3] = 0.18 μM, [Profilin] = 6 μM, and [Phalloidin] = 12 μM. Samples are labeled at a labeling ratio of R = 0.05.

\subsection*{Hencky strain.}
The Hencky strain $\varepsilon$ is given by \cite{reesBasicEngineeringPlasticity2012}

\begin{equation*}
    \varepsilon = \int \frac{dl}{l}
\end{equation*}

where the integral is performed over a sequence of incremental changes in line elements $dl$ with respect to their lengths $l$. This quantity is preferred over the engineering strain for large deformations, typically above 1\% strain, since for deformations larger than 1\% Hencky strain is path-independent\cite{reesBasicEngineeringPlasticity2012}. By taking length $l = \sqrt{A}$ as the square root of the imaged area of the gel, and discretizing $dl \approx \Delta l$, we have:

\begin{equation*}
E_H^{\text{bnd}}(t)=\varepsilon = \int \frac{dl}{l} \approx \sum \frac{\Delta l}{l_i} = \sum
\frac{l_{i} - l_{i+1}}{l_i} = \sum
\frac{\sqrt{A_i} - \sqrt{A_{i+1}}}{\sqrt{A_i}}
\end{equation*}

\subsection*{Tensorial Strain and Hencky Strain from PIV}

\subsubsection*{Displacement Field and Deformation Gradient}
To quantify the spatiotemporal deformation of the actomyosin networks during contraction, we extracted displacement and velocity fields using particle image velocimetry (PIV) implemented in MATLAB. Time-lapse fluorescence microscopy images of the actin network were analyzed using the open-source MATLAB PIVlab framework~\cite{thielicke_particle_2021}, which computes local displacements by cross-correlating interrogation windows between successive image frames.

Particle Image Velocimetry (PIV) provides the displacement field 
\[
u(X,t) = 
\begin{bmatrix}
u_x(X,t) \\[4pt]
u_y(X,t)
\end{bmatrix},
\qquad X = (x,y).
\]

From this field, we compute the deformation gradient tensor
\begin{equation}
F(X,t) = I + \nabla u(X,t)
=
\begin{bmatrix}
1 + \dfrac{\partial u_x}{\partial x} & \dfrac{\partial u_x}{\partial y} \\[8pt]
\dfrac{\partial u_y}{\partial x} & 1 + \dfrac{\partial u_y}{\partial y}
\end{bmatrix}.
\end{equation}
The tensor \(F\) describes the local mapping of infinitesimal material elements from the reference to the deformed configuration.

\subsubsection*{Finite Strain Measure: Green--Lagrange Tensor}

To accurately capture finite deformation during gel contraction, we compute the nonlinear Green--Lagrange strain tensor:
\begin{equation}
\mathbf{E}(X,t)
= \frac{1}{2}\left(F^\mathsf{T} F - I\right).
\end{equation}

The in-plane components used in our analysis are
\[
E_{xx}(X,t), \qquad E_{yy}(X,t), \qquad E_{xy}(X,t).
\]

\subsubsection*{Local Areal Strain and Hencky Strain}

We define the in-plane normal strain components as
\[
\epsilon_{xx}(X,t) = E_{xx}(X,t), 
\qquad 
\epsilon_{yy}(X,t) = E_{yy}(X,t).
\]

A local measure of in-plane dilatation is obtained from the sum of normal components:
\begin{equation}
\mathrm{div}\,u(X,t) = \epsilon_{xx}(X,t) + \epsilon_{yy}(X,t).
\end{equation}

This quantity approximates the local fractional change in area at each spatial location.

The corresponding local Hencky (logarithmic) strain is then defined as
\begin{equation}
E_H(X,t) = \frac{1}{2}\,\ln\!\left( 1 + \mathrm{div}\,u(X,t) \right).
\end{equation}

This form is consistent with the Hencky strain definition based on an areal stretch ratio
\[
\lambda_A(X,t) \approx 1 + \mathrm{div}\,u(X,t),
\qquad 
E_H = \frac{1}{2}\ln\lambda_A.
\]

\subsubsection*{Spatial Averaging and Temporal Contractility}

To summarize the local Hencky strain field into a global tensorial measure, we take the spatial average:
\begin{equation}
E_H^{\text{ten}}(t)
= \frac{1}{A} 
\int_{\Omega} E_H(X,t)\,dA,
\end{equation}
where \(\Omega\) is the gel domain.

The cumulative Hencky strain over time is computed as
\begin{equation}
E_H^{\text{cum}}(t_k) 
= \sum_{i=1}^{k} E_H^{\text{ten}}(t_i).
\end{equation}

For comparison, we also compute a linearized contractility measure based on normal strain components:
\begin{equation}
C(t) = \frac{1}{A}
\int_{\Omega} \left( \epsilon_{xx}(X,t) + \epsilon_{yy}(X,t) \right) dA.
\end{equation}

The tensorial Hencky strain \(E_H^{\text{ten}}(t)\) and boundary Hencky strain 
\(E_H^{\text{bnd}}(t)\) are directly compared to quantify agreement between 
local (PIV-derived) and global (outline-derived) deformation measures. For main text we have kept tensorial Hencky strain as a strain measure as it gives us spatiotemporal information of a strain.

\begin{figure}
    \centering
    \includegraphics[width=\linewidth]{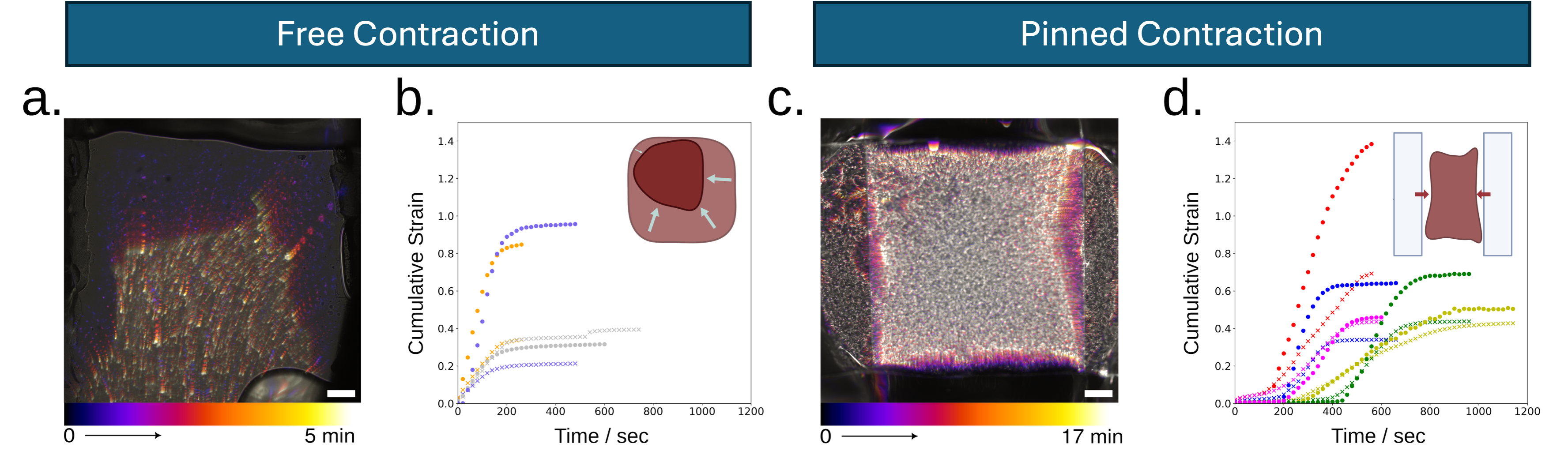}
\caption{Graphical representations of the two contraction regimes considered for analysis (a) free contraction - our control population (N=3), (b)Cumulative Hencky strain for N=3 control samples (c) pinned contraction (N-5). (d)  Cumulative Hencky strain for N=5 pinned samples with a detach relaxation pathway. For panels b, and d data is shown for tensorial Hencky strain  $E_H^{\text{ten}}(t)$ determined using PIV (x's), and for boundary Hencky strain $E_H^{\text{bnd}}(t)$ determined using area-based (circles) methodologies. }
    \label{sfig_strain}
\end{figure}

\subsection*{Image processing}

Images were preprocessed by taking only Red channel and then normalizing the intensity. To post process data output displacements from PIV were saved as .csv file per frame and then MATLAB was used to get properties as per above mentioned formulas.

For image processing, all frames were preprocessed using local adaptive thresholding from skimage filters of the Scikit-image library. The Segment Anything Model (SAM) from Meta AI is used to create masks of the actomyosin gel as it contracts of all frames imaged. Using the SamPredictor, points (x, y), which can be labeled as foreground or background points, and boxes (x1, y1, x2, y2), are entered to specify the region of analysis that yields the best mask. The quality of the mask is assessed by how closely the mask aligns with the actual gel. The mask is then cleaned, removing inaccurate holes or discontinuities as needed. This mask, stored as an array of "True" and "False" values, can then be used to find the area of the gel, enabling Hencky strain calculations. Note that the ViT-H SAM model was used that the gel was masked between the walls only. 

\subsection*{Sample preparation - pulsed}
Actin was purified from rabbit psoas skeletal muscle from Pel-Freeze using a GE Superdex 200 Increase HiScale 16/40 column and stored at $\SI{-80}{\degreeCelsius}$ in G-Buffer ($\qty{2}{\milli M}$  tris-hydrochloride pH 8.0, $\qty{0.2}{\milli M}$ disodium adenosine triphosphate (ATP), $\qty{0.2}{\milli M}$ calcium chloride, and $\qty{0.2}{\milli M}$ dithiothreitol). All protein stocks were clarified of aggregated proteins at 100 000 gg for five minutes upon thawing and used within seven days. The G-actin concentration in the supernatant was determined by measuring the solution absorbance at $\qty{290}{\nano\meter}$ with a NanodDrop 2000 (Thermo Scientific, Wilmington, DE, USA) and using an extinction coefficient of $\qty{26600}{M^{-1}\centi\meter^{-1}}$. Actin is dialyzed in monomeric buffer containing \qty{0.001}{mM} ATP concentration to prevent large pre-contraction events due to residual ATP present in the standard actin monomeric storage buffer.

\subsection*{Pulsed assay formulation}
In addition to the standard contractile assay reagents (else ATP), pulsed excitation experiments utilize NPE-caged ATP (CATP), and phalloidin. Assays are formulated with $\left[\mathrm{CATP}\right] = \qty{0.1}{m M}$, and $[\mathrm{Phalloidin}] = 12\mu M$. Note that mixed assay are added to evaporated phalloidin as the final mixing step before loading into the chamber. Phalloidin is stored in methanol which will destroy the actin network if not sufficiently evaporated using a clean nitrogen line. Samples are labeled at a labeling ratio of $R=0.05$.
\subsection*{Cytoskeleton Composite Methods Section}

All cytoskeleton proteins are purchased from Cytoskeleton: rabbit skeletal actin (Cytoskeleton AKL99), HiLyte Fluor-555-labeled actin (Cytoskeleton AR07), porcine brain tubulin (Cytoskeleton T240), HiLyte Fluor-647-labeled tubulin (Cytoskeleton TL670M), and rabbit skeletal myosin II (Cytoskeleton MY02). Actin is reconstituted to 2 mg/mL in 2.0 mM Tris (pH 8), 0.2 mM ATP, 0.5 mM DTT, 0.1 mM CaCl2. Tubulin is reconstituted to 5 mg/mL in PEM-100 [100 mM PIPES (pH 6.8), 2 mM MgCl2, and 2 mM EGTA]. Myosin II is reconstituted to 10 mg/ml in 25 mM PIPES (pH 7.0), 1.25 M KCl, 2.5\% sucrose, 0.5\% dextran, and 1 mM DTT. Following reconstitution, all proteins are flash frozen and stored at -80 C in single-use aliquots. 

Following previously published protocols (https://www.science.org/doi/full/10.1126/sciadv.abe4334, https://www.jove.com/v/64228/reconstituting-characterizing-actin-microtubule-composites-with)
, to prepare actin-driven networks of entangled actin or co-entangled actin and microtubules, we mix together  4 – 5.8 µM actin monomers, at 1:5 ratio of labeled:unlabeled units, and 0 – 1.74 µM tubulin dimers, at 1:10 ratio of labeled:unlabeled units, in PEM-100 supplemented with 0.1\% Tween20, 2 mM ATP and 2 mM GTP. We add phalloidin (Apollo Scientific BIP0842) at a 1:1 ratio with actin monomers, and 5 µM paclitaxel (Taxol, Sigma-Aldrich) to stabilize actin filaments and microtubules, respectively. Following incubation at 37 C for 1 hour, we add 10 µM photo-sensitive myosin inhibitor, blebbistatin (Sigma Aldrich B0560), 2.5 – 8 µM myosin, and an oxygen-scavenging system [45 μg/mL glucose, 0.005\% β-mercaptoethanol, 43 μg/mL glucose oxidase, 7 μg/mL catalase, 2 mM Trolox (Sigma)] to inhibit photobleaching.

For all experiments, we load samples into rectangular glass capillary tubes (VitroTubes #5012-100) with 0.10 x 2.00 mm inner dimensions cross-sectional area, 0.1 mm thickness and X mm length, which we adhered to a glass microscope slide. To prevent adsorption of proteins to capillary walls, we fill chambers with 10 mg/mL solution of BSA (bovine serum albumin) in 1\% Tween, incubate for 5 mins, then flushed out with compressed air. We then flow in the polymerized samples into the capillaries via capillary action and seal the open ends with UV-curable epoxy.

We performed all experiments using a Nikon A1R laser scanning confocal microscope with a 4x 2.0 NA objective (Nikon), with a 561-nm laser and 565/591-nm excitation/emission filters to image actin and 640-nm laser and 624±20/692±20-nm excitation/emission filters to image tubulin. For composites with blebbistatin, we also illuminated the field-of-view with 405-nm laser to locally deactivate blebbistatin to enable myosin activity. To capture the entire length of the capillary, we acquired two-channel time-series of X x X square-pixel (X µm x X µm) images by sequentially capturing eight 512 x 512 images (3181 μm x 3181 μm) at increasing vertical distances with 10\% overlap with the previous image, to span a total vertical distance of XX µm. Following acquisition, we stitched images together using XXX software, resulting in a single time-series of dimensions XXX. For samples containing blebbistatin, we captured one stitched together frame every 15 (Fig 5h, iii) or 20 (Fig 5h, ii) seconds for 3 hours during 405-nm illumination, followed by every 60 s for 12 hours (x total frames). For samples without blebbistatin (Fig 5h, i, iv) we captured 1 stitched frame every 60 seconds for 10 minutes (x total frames).

\subsection*{Wave speed.}
For the case of local rupture, we perform preliminary analysis of strain wave speed in the gel. We are limited in the extent by which we can draw conclusions due to the small number of observations of this effect (N=1). 

\begin{figure}
    \centering
    \includegraphics[width=\linewidth]{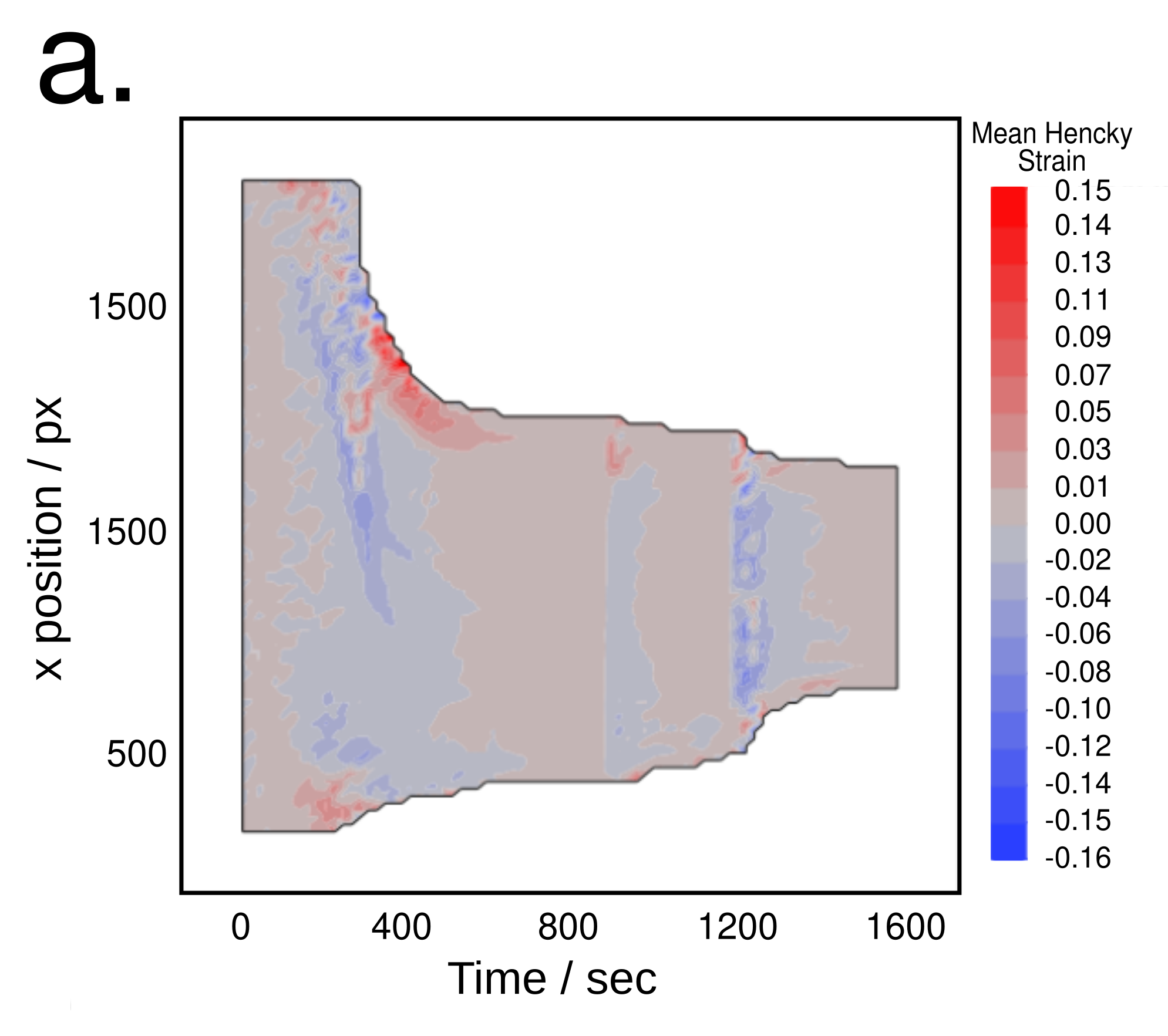}
\caption{(a) Kymograph showing average strain change with time. Inhomogeneous strain behavior shows the presence of contractile front followed by tensile strain region moving through material.}
    \label{sfig_wavespeed}
\end{figure}

\subsection*{Simulation}

\subsubsection*{Introduction}
We model active contraction and depinning in an actomyosin network by modeling the actin network using a triangular lattice. Fascin crosslinkers are placed at locations where two or more filaments cross. Active contractions by myosin motors is implemented using force dipoles which have the effect of reducing the rest lengths of the edges of the network.

Our initial condition is a regular two-dimensional lattice where the positions of the filaments and crosslinkers are well-defined. This arrangement reflects the high degree of crosslinking in the actomyosin gel at the beginning of the experiment. The relevant length scale, i.e., the distance between crosslinks is about an order of magnitude smaller than the persistence length of actin filaments\cite{Howard2001-pz}, so thermal fluctuations that lead to filament bending can be ignored. The high degree of network crosslinking also makes drift of the filaments through  the fluid and steric interactions with neighboring filaments less likely, so we also ignore repulsive forces between filaments at short distances.

The framework of modeling the mechanics of actin networks using a diluted triangular lattice and myosin activity through force dipoles has been used in earlier studies~\cite{broedersz2011molecular, kumar2023range, lee2021active}. While these studies model the evolution of the network using energy minimization, we evolve our system by solving the equations of motion of the nodes of the network in an overdamped limit.

\subsubsection*{Network construction and dynamics}
\label{sec:network}
We model the actin network as a diluted two-dimensional triangular lattice with a lattice spacing $\ell_0$. Each possible edge on the lattice is placed with a probability $p_{\text{occ}}$, which we chose to be $0.8$. A collinear sequence segments constitutes an actin filament. Due to the random initial placement of edges, we have filaments of variable lengths. Initially, filaments are oriented at 0, 60, or 120 degree angle with the positive x-axis. A filament is represented as a series of nodes separated by $\ell_0$. The edges joining the nodes are modeled as springs that resist changes to their lengths. Each hinge formed at the intersection of two internal edges also resists angular deformation away from the resting angle of 180 degrees.

If two filaments cross at a node on the lattice, a passive crosslinker is placed at that location. If three filaments cross at a node, a crosslinker is placed joining two randomly picked filaments; the third filament can move without constraints through the node, as a "phantom" filament\cite{broedersz2011molecular}. Nodes at the crosslinkers belong to both the linked filaments; thus their motion is governed by the forces coming from both the filaments. In this model, crosslinkers are placed at the beginning of the simulation and they do not unbind.

Active contraction in the system through myosin motors is implemented by modeling the motors as force dipoles. When a force diploe is randomly placed on an edge, it reduces rest length of the edge by an amount $\delta$. The force dipole immediately detaches, reflecting the low duty-ratio of myosin motors\red{cite}, and are placed at the same edge or a different edge at a later time point randomly. The reductions in rest length, however, accumulate over time. For more details on the motor concentration, and the timescales of attachment and detachment, see Section \ref{sec:units}. 

The dynamics of the system is implemented by solving the equations of motion of the beads in the overdamped limit. Specifically, bending and stretching forces on the nodes are considered. The stretching or contraction of an edge is penalized by a stretching energy governed by the stretching modulus $k_{\text{stretch}}$ and an anharmonicity factor $\alpha$. Three consecutive beads on a filament constitute a hinge; deviation of the hinge angle from \SI{180}{\degree} is penalized by a bending energy governed by a bending modulus $k_{\text{bend}}$. The equation of motion for the nodes can be written as

\begin{align}
 \dot{\vb{x}_i} &= \sum_i \left( -\frac{\nabla_i U_{\text{stretch}}}{\gamma_{\text{stretch}}} - \frac{\nabla_i U_{\text{bend}}}{\gamma_{\text{bend}}} \right) \label{eqn:eqn_of_motion} \\ 
 U_{\text{stretch}} &= \sum_{\left<ij\right>} p_{ij} k_{\text{stretch}} \left( \frac{1}{2} \left(r_{ij} - r_{ij}^0 \right)^2 + \alpha \left(r_{ij} - r_{ij}^0 \right)^4 \right) \label{eqn:stretch_energy} \\
 U_{\text{bend}} &= \sum_{\left<ijk\right>} p_{ij} p_{jk} \frac{k_{\text{bend}}}{2} \left( \theta_{ijk} - \pi \right)^2  \label{eqn:bend_energy}
\end{align}

In the above, $r_{ij}$ is the length of the edge connecting two nearest neighboring nodes $i$ and $j$; $r_{ij}^0$ is its rest length, which is taken to be $\ell_0$ at the beginning of the simulation but can be reduced in steps of $\delta$ by the action of myosin motors. $\theta_{ijk}$ is the angle at the node $j$ that forms the hinge between edges $ij$ and $jk$. $p_{ij}$ and $p_{jk}$ are Boolean variables that indicate whether an edge between nodes $i$ and $j$ exists or not, by assuming values 1 or 0 respectively.

The forces computed in Equation \ref{eqn:eqn_of_motion} are related to the node velocities through the viscous drag terms $\gamma_{\text{stretch}}$ and $\gamma_{\text{bend}}$. For the geometry of the edges of the network, treating the edges as cylinders with length $L$ and radius of cross section $r$ is a natural choice. $\gamma_{\text{stretch}}$ is the drag when the cylinder moves parallel to its central axis, and $\gamma_{\text{bend}}$ is the drag when the cylinder moves perpendicular to its central axis. These values are given by \cite{Howard2001-pz}

\begin{align}
 \gamma_{\text{stretch}} = \gamma_{\parallel} &= \frac{2\pi \eta L}{\ln(L/2r) - 0.2} \label{eqn:parallel_drag} \\
 \gamma_{\text{bend}} = \gamma_{\perp} &= \frac{4\pi \eta L}{\ln(L/2r) + 0.84} \label{eqn:perp_drag}
\end{align}
where $\eta$ is the dynamic viscosity of the fluid.

While simulating pinned systems, the nodes at the boundaries are held fixed until the force on them exceeds a threshold force $f_{\text{Depin}}$. Once the force on a node exceeds the threshold force, the node detaches from the boundary and moves according to the force on it. For the control case of free contraction, all beads are free to move from the beginning of the simulation. 

\subsubsection*{Units}
\label{sec:units}
Our system is non-dimensionalized by three physical quantities: lattice spacing $\ell_0$, viscosity of the fluid $\eta$, and the bending modulus of the filaments $k_{\text{bend}}$. All other quantities can be expressed as products of powers of these three quantities. For example, the unit of time $t = \eta\ell_0^3/k_{\text{bend}}$. 

The lattice spacing $\ell_0$ is taken to correspond to \SI{0.27}{\micro\metre}, or 100 times the radius of a G-actin monomer~\cite{bionumFilamentSize}. $\ell_0$ is also the separation between two crosslinkers on a filament. We have separately simulated systems with a smaller lattice spacing. Given the scaling of the non-dimensionalized time on the lattice spacing $(t \sim \ell_0^4)$, this necessitates a significantly smaller integration time step. However, we find no meaningful difference in the evolution of the network. \red{Need to back this up with a figure?}. Hence, we report all our results using $\ell_0 = \SI{0.27}{\micro\metre}$.

The experimental value of bending rigidity of f-actin $\kappa$ is \SI{60e-27}{\newton\metre^2}~\cite{Howard2001-pz}. The bending modulus is given by
$k_{\text{bend}} = \frac{\kappa}{2\ell} = \frac{\num{60e-27}}{2\times \num{0.27e-6}} \si{\newton\metre} = \SI{1.1e-19}{\newton\metre} = \SI{110}{\pico\newton\nano\metre} $

\red{Do we need some explanation or schematic deriving the $\kappa/2\ell$ relation?}

For a cylindrical rod with radius of cross section $r$, the stretching modulus $k_{\text{stretch}}$ and the bending modulus $k_{\text{bend}}$ are related by $k_{\text{stretch}} = 8k_{\text{bend}}/r^2$.

To obtain an integration time step that is feasible to simulate the network for tens of minutes and to have numerical stability, we reduce the model stretching modulus by a factor of 100 and choose the viscosity of the fluid to be $\num{1e3}$ times the viscosity of water. As a result, our unit of time corresponds to \SI{0.12}{\second} of physical time. See Table \ref{tab:sim_param}.

A motor reduces the rest length of the network edge that it is acting on, by an amount $\delta$. Our value of $\delta = \SI{5}{\nano\metre} = 0.0185\ell$ is experimentally informed~\cite{stam2015isoforms}. As motors consume ATP, the concentration of free motors decays with time. We assume a first-order decay of the number of active motors, given by the decay time $\tau_d$ as $N_m = N_m^0 \exp(-t/\tau_d)$. $N_m^0$ is taken to be $0.1$ times the number of crosslinkers.

The values of the simulation parameters in both SI and non-dimensionalized units is given in Table \ref{tab:sim_param}.

\begin{table}
\centering
\caption{Simulation parameters and their values in SI and non-dimensionalized units}
\begin{tabular}{|p{2.0in}|p{1.5in}|p{1.5in}|p{1.0in}|} \hline
 \textbf{Quantity} & \textbf{Symbol and expression} & \textbf{Value in SI units} & \textbf{Value in non-dimensionalized units} \\ \hline 
 Lattice spacing & $\ell$ & \SI{270}{\nano\metre} & $1$ \\ \hline
 Bending modulus & $k_{\text{bend}} = \kappa/2\ell$ & \SI{110}{\pico\newton\nano\metre} & $1$ \\ \hline
 Fluid viscosity & $10^3\dag \times \eta$ & $10^3 \times \SI{0.69e-3}{\pascal\second}$ & $1$ \\ \hline
 Time (unit) & $t = \eta\ell_0^3/k_{\text{bend}}$ & \SI{0.12}{\second} & $1$ \\ \hline
 Radius of cross section & $r = \ell/100$ & \SI{2.7}{\nano\metre} & $0.01$ \\ \hline
 Stretching modulus & $k_{\text{stretch}} = 10^{-2}\dag \times 8k_{\text{bend}}/r^2$ & $10^{-2} \times \SI{120}{\pico\newton\per\nano\metre} $ & $800$ \\ \hline
 Spring anharmonicity & $\alpha$ & & $0.025$ \\ \hline
 Stretch (parallel) drag coeff & $\gamma_{\parallel}$, see Equation \ref{eqn:parallel_drag} & \SI{3.153e-4}{\pico\newton\second\per\nano\metre} & $1.69$  \\ \hline
 Bend (perpendicular) drag coeff & $\gamma_{\perp}$, see Equation \ref{eqn:perp_drag} & \SI{4.926e-4}{\pico\newton\second\per\nano\metre} & $2.64$  \\ \hline
 Motor step length & $\delta$ & \SI{5}{\nano\metre} & $0.0185$ \\ \hline
 Horizontal depinning force (per node) &$f_{\text{Depin-h}} = k_{\text{stretch}}\cdot n\delta$, $n=10-20$ & $60 - 120$ \si{\pico\newton}& $148 - 296$ \\ \hline
 Vertical depinning force (per node) &$f_{\text{Depin-v}} = k_{\text{stretch}}\cdot m\delta$, $m=5$ & $30$ \si{\pico\newton} & $74$ \\ \hline
 Motor pull probability rate & $k_{\text{on}}$ & \SI{0.2}{\per\second} & $0.024$ \\ \hline 
 Myosin decay time constant & $\tau_d$ & \SI{120}{\second} & $1000$ \\ \hline
 Lattice occupation probability & $p_{\text{occ}}$ & - & $0.8$ \\ \hline
\end{tabular}
\label{tab:sim_param}
$\dag$ Factors of $10^3$ and $10^{-2}$ are included to obtain a computationally feasible integration time step, and to maintain numerical stability, respectively.
\end{table}

\subsubsection*{Kurtosis can show presence of intermittent dynamics}

To investigate the presence of intermittent dynamics in stress-bearing gels, we analyze the spatiotemporal evolution of strain fields derived from displacement data. The displacement field, denoted as $\mathbf{u}(x,y,t)$, is used to compute the strain field:

\begin{equation}
\epsilon(x,y,t) = \nabla \mathbf{u}(x,y,t)
\end{equation}

From this, we extract strain components such as $\epsilon_{xx}, \epsilon_{yy}, \epsilon_{xy}$ depending on the context.

To quantify temporal changes, we compute the temporal autocorrelation matrix $\mathbf{C}$, where each element $C_{ij}$ represents the normalized correlation between strain fields at times $t_i$ and $t_j$:

\begin{equation}
C_{ij} = \frac{\sum_{x,y}\left[\epsilon(x,y,t_i)-\bar{\epsilon}(t_i)\right]\left[\epsilon(x,y,t_j)-\bar{\epsilon}(t_j)\right]}{\sqrt{\sum_{x,y}\left[\epsilon(x,y,t_i)-\bar{\epsilon}(t_i)\right]^2}\sqrt{\sum_{x,y}\left[\epsilon(x,y,t_j)-\bar{\epsilon}(t_j)\right]^2}}
\end{equation}

Here, $\bar{\epsilon}(t)$ is the spatial average of strain at time $t$:

\begin{equation}
\bar{\epsilon}(t) = \frac{1}{N_x N_y} \sum_{x,y} \epsilon(x,y,t)
\end{equation}

To further characterize the distribution of correlations, we compute the kurtosis $\kappa(t_i)$ for each row of the autocorrelation matrix:

\begin{equation}
\kappa(t_i) = \frac{\frac{1}{N}\sum_{j=1}^{N}(C_{ij}-\mu_i)^4}{\sigma_i^4}
\end{equation}

where

\begin{equation}
\mu_i = \frac{1}{N}\sum_{j=1}^{N} C_{ij}, \quad \sigma_i = \sqrt{\frac{1}{N}\sum_{j=1}^{N}(C_{ij}-\mu_i)^2}
\end{equation}

High kurtosis values indicate the presence of outliers in the correlation distribution, which may signal intermittent or abrupt changes in the strain field.

\section{Additional results}
\subsection*{Strain Field.}
\begin{figure}
    \centering
    \includegraphics[width=\linewidth]{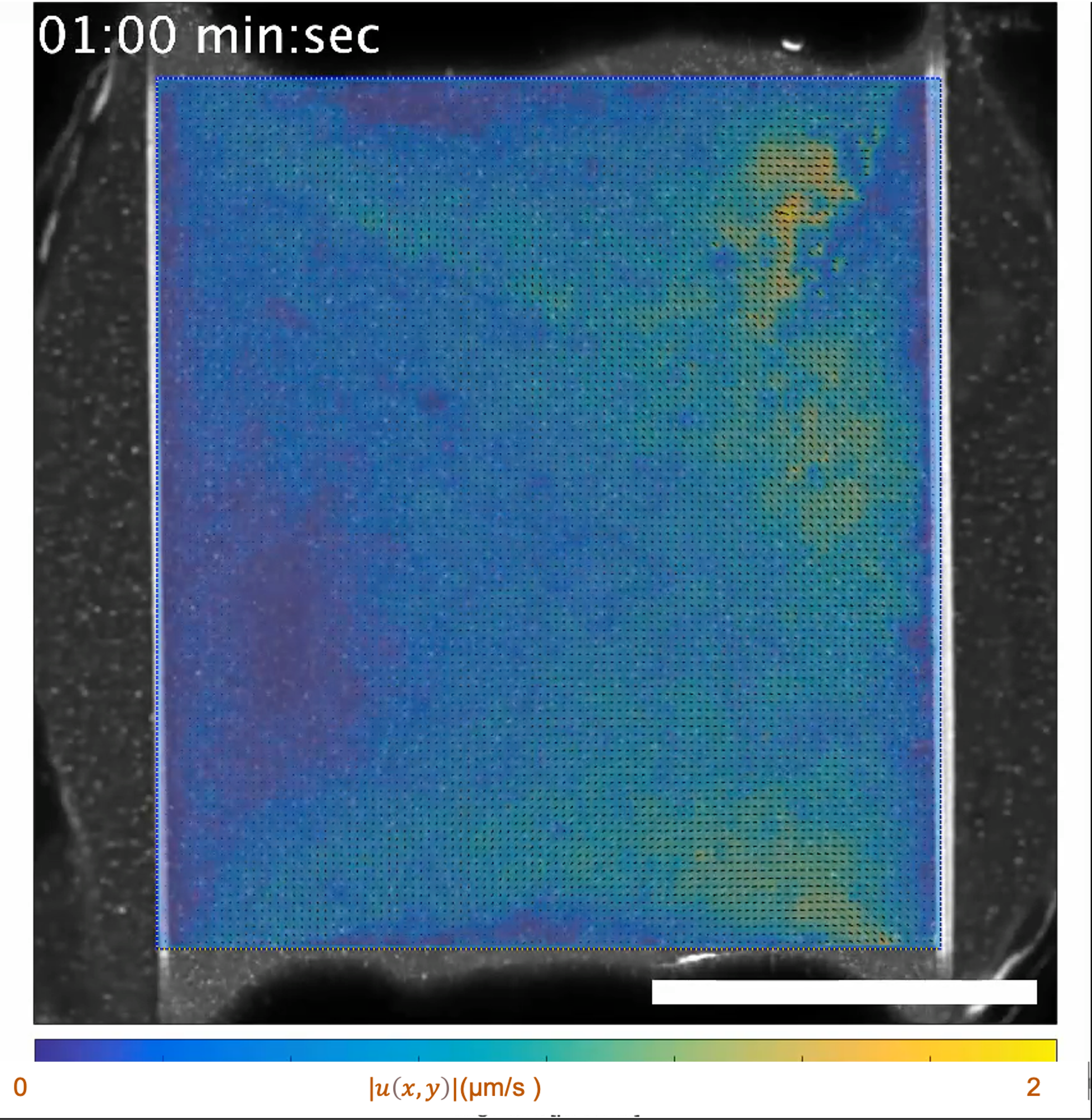}
\caption{Snapshot of Pinned experiment showing spatiotemporal variation in displacement when no global change in area}
    \label{sfig_Disp}
\end{figure}

To characterize the mechanical state of the system prior to the onset of bulk contraction, we analyze instantaneous strain fields extracted during the lag phase, defined as the interval between ATP addition and the emergence of a measurable macroscopic strain rate. During this phase, the network exhibits negligible global deformation while still supporting localized stress buildup and heterogeneous microstructural rearrangements.
During the lag phase, the strain field is characterized by spatially intermittent, low-amplitude fluctuations rather than coherent deformation modes as shown in Fig.(\ref{sfig_Disp}).. These fluctuations are not uniformly distributed but instead appear as mesoscopic patches of positive and negative strain, reflecting local stress accumulation, filament rearrangements, and transient bond remodeling. Importantly, despite the absence of a net macroscopic strain rate, the spatial variance of the strain field is non-zero and can serve as an early indicator of impending mechanical instability.

\subsection*{Rise Time.}
\begin{figure}
    \centering
    \includegraphics[width=\linewidth]{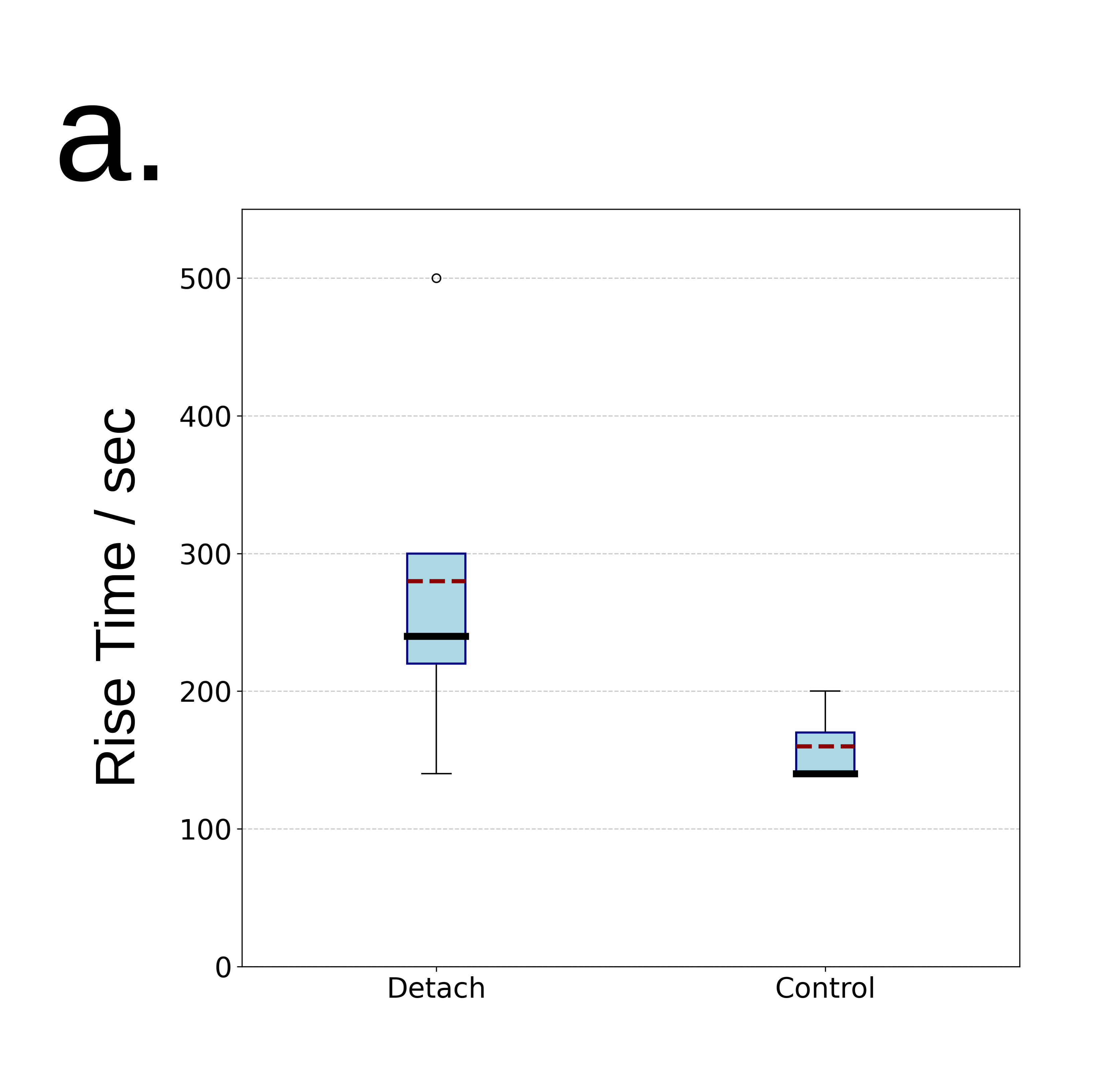}
\caption{(a) Rise Time distributions comparing the time it takes to contract between pinned, and control datasets. (p=0.121)}
    \label{sfig_risetime}
\end{figure}

We also compare the dynamics of contraction via the Rise Time of each sample, shown in Fig.(\ref{sfig_risetime}a). This is determined by subtracting the time it takes the sample to reach 10\% maximum strain value from the time it takes to reach 90\% maximum strain value. We find a lengthening of the Rise Time values associated with pinned contraction (p=0.121). We believe that this slow down is likely related to the available ATP for the gel, once it begins to contract freely after detaching from the boundary. Given that an amount of ATP was consumed prior to detaching from the wall, the gel would have less ATP available, which in turn would result in less processivity among myosin II motors which we believe results in a slower contraction rise. An alternative reasoning is that there is some form of stiffening due to the pinning of the gel. Forces sustained at the boundary are likely to increase tension within the gel which can alter the velocity of myosin motors. Further, it is well known that actin filaments under tension stiffen \cite{koenderinkActiveBiopolymerNetwork2009,stormNonlinearElasticityBiological2005}. In order to test this and decouple the two effects, further experiments would need to be performed. We note that this result varies from our previous findings in which the rise time of freely contracting gels with constant ATP concentrations was ~\qty{250}{\second}.

\subsection*{Stall Time.}
\begin{figure}
    \centering
    \includegraphics[width=\linewidth]{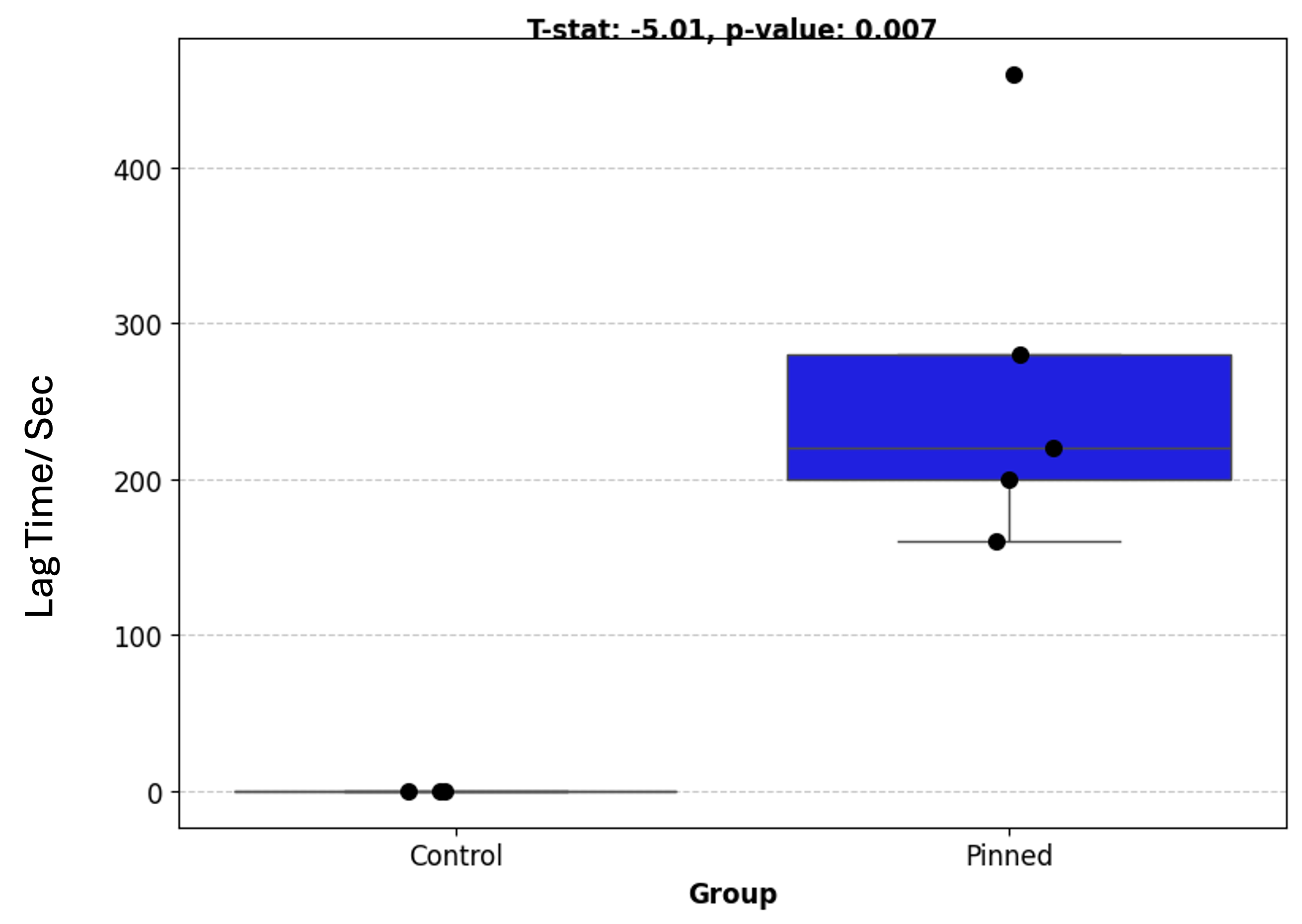}
\caption{(a) Stall Time distributions comparing the time it takes to start contraction phase between pinned, and control datasets. (p=0.007)}
    \label{sfig_lagtime}
\end{figure}

To quantify the dynamical delay between the onset of activity and the mechanical response of the network, we define a lag time based on the temporal evolution of the cumulative strain field for both control and pinned conditions.
The stall time, $\tau_{lag}$	
 , is defined as the time interval between the onset of activity (taken as t=0, corresponding to ATP addition or initiation of contraction) and the first detectable systematic deviation of the strain signal from the baseline noise floor. Practically, we determine this by identifying the time at which $\varepsilon(t)$ exceeds a threshold set by the pre-activation fluctuations (mean + 2–3 standard deviations of the baseline signal). 
 The extracted $\tau_{lag}$ values are reported in Fig.(\ref{sfig_lagtime}).

\bibliographystyle{unsrt}
\bibliography{biblio}